\documentclass[fleqn,usenatbib]{mnras}

\usepackage{newtxtext,newtxmath}

\usepackage[T1]{fontenc}

\DeclareRobustCommand{\VAN}[3]{#2}
\let\VANthebibliography\thebibliography
\def\thebibliography{\DeclareRobustCommand{\VAN}[3]{##3}\VANthebibliography}

\usepackage{graphicx}	
\usepackage{amsmath}	
\usepackage{gensymb}
\usepackage{multirow,booktabs}
\usepackage{scalerel}

\newcommand{\CIV}{\ion{C}{iv}}
\newcommand{\CIVdist}{\ion{C}{iv} distance}
\newcommand{\HeII}{\ion{He}{ii}}
\newcommand{\aox}{$\alpha_\text{ox}$}
\newcommand{\aouv}{$\alpha_\text{ouv}$}
\newcommand{\auvx}{$\alpha_\text{uvx}$}
\newcommand{\kms}{km\,s$^{-1}$}
\newcommand{\Lx}{$L_{2\,\text{keV}}$}
\newcommand{\BHM}{$M_\text{BH}$}
\newcommand{\LLEdd}{$L/L_\text{Edd}$}
\newcommand{\HI}{H\protect\scaleto{$I$}{1.2ex}}

\usepackage{orcidlink}

\defcitealias{rankine_bal_2020}{R20}
\defcitealias{rankine_placing_2021}{R21}
\defcitealias{shlentsova_x-ray_2026}{S26}

\usepackage{soul}
\newcommand{\hli}[1]{#1}
\newcommand{\hlii}[1]{#1}
\newcommand{\hliii}[1]{#1}

\title[X-ray selected C\,{\normalsize \textit{IV}} winds]{C\,{\Large \textbf{IV}} wind properties of the SDSS-V X-ray selected quasars: strong optical-to-UV emission is key regardless of X-ray strength}

\author[A. L. Rankine et al.]{%
Amy L. Rankine$^{\orcidlink{0000-0002-2091-1966}}$,$^{1}$\thanks{E-mail: amy.rankine@ed.ac.uk (ALR)}
David Homan$^{\orcidlink{0000-0002-3243-874X}}$,$^{2,3}$
James Aird$^{\orcidlink{0000-0003-1908-8463}}$,$^{1}$
Pranavi Hiremath$^{\orcidlink{0009-0005-0072-6973}}$,$^{1}$
Scott F. Anderson$^{\orcidlink{0000-0002-6404-9562}}$,$^{4}$
\newauthor
Roberto J. Assef$^{\orcidlink{0000-0002-9508-3667}}$,$^{5}$
Franz E. Bauer$^{\orcidlink{0000-0002-8686-8737}}$,$^{6}$
W. N. Brandt$^{\orcidlink{0000-0002-0167-2453}}$,$^{7, 8, 9}$
Marcella Brusa$^{\orcidlink{0000-0002-5059-6848}}$,$^{10, 11, 12}$
Johannes Buchner$^{\orcidlink{0000-0003-0426-6634}}$,$^{13}$
\newauthor
Maria Chira$^{\orcidlink{0000-0003-3116-9258}}$,$^{14}$
Yaherlyn D\'iaz$^{\orcidlink{0000-0002-8604-1158}}$,$^{15}$
Patrick B. Hall$^{\orcidlink{0000-0002-1763-5825}}$,$^{16}$
Anton M. Koekemoer${\orcidlink{0000-0002-6610-2048}}$,$^{17}$ 
Mirko Krumpe,$^2$
\newauthor
Georg Lamer$^{\orcidlink{0000-0001-6238-2321}}$,$^{2}$
Teng Liu$^{\orcidlink{0000-0002-2941-6734}}$,$^{18, 19}$
Sean Morrison$^{\orcidlink{0000-0002-6770-2627}}$,$^{20}$
Blessing Musiimenta$^{\orcidlink{0000-0001-8162-0719}}$,$^{10, 11, 12}$
C. A. Negrete$^{\orcidlink{0000-0002-1656-827X}}$,$^{21}$
\newauthor
Qingling Ni$^{\orcidlink{0000-0002-8577-2717}}$,$^{13}$
Paola Rodr\'iguez Hidalgo$^{\orcidlink{0000-0003-0677-785X}}$,$^{22}$
Mara Salvato$^{\orcidlink{0000-0001-7116-9303}}$,$^{13}$
Donald P. Schneider$^{\orcidlink{0000-0001-7240-7449}}$,$^{7, 8}$
Yue Shen${\orcidlink{0000-0003-1659-7035}}$,$^{20,23}$
\newauthor
Matthew J. Temple$^{\orcidlink{0000-0001-8433-550X}}$,$^{24}$
Dus\'an Tub\'in-Arenas$^{\orcidlink{0000-0002-2688-7960}}$,$^2$
Dominika Wylezalek$^{\orcidlink{0000-0003-2212-6045}}$$^{25}$
\\
Affiliations are listed at the end of the article\\
}

\date{Accepted XXX. Received YYY; in original form ZZZ}

\pubyear{2026}

\begin{document}
\label{firstpage}
\pagerange{\pageref{firstpage}--\pageref{lastpage}}
\maketitle

\begin{abstract}

We present an investigation of the rest-frame optical/UV and X-ray properties for a sample of 3027 X-ray selected quasars between $1.5 \leq z \leq 3.5$ detected in the deepest Spectrum Roentgen Gamma/\textit{eROSITA} data available and observed by the fifth iteration of the Sloan Digital Sky Survey (SDSS-V). We parametrize the {\CIV}\,$\lambda1549$ emission line to infer the strength of accretion disc winds and perform X-ray spectral fitting. The X-ray spectral properties -- namely, the 2\,keV monochromatic luminosity (\Lx) and spectral slope -- are not strongly correlated with wind strength. Despite this result, the X-ray selected sample is shifted towards lower {\CIV} blueshifts and higher equivalent widths than the optically selected sample observed in previous SDSS surveys, and matching in optical luminosity, redshift, and Eddington ratio does not reduce these differences. We estimate the far-UV luminosity using the {\HeII}\,$\lambda1640$ line luminosity and define the slopes between this and the 2500\,{\AA} monochromatic luminosity ($L_{2500}$) and {\Lx} ({\aouv} and {\auvx}, respectively) in a similar manner to the familiar {\aox} parameter, which tracks the spectral slope between $L_{2500}$ and {\Lx}. The quantity {\aouv} is more strongly correlated with wind strength in our sample than {\aox}. We show that the correlation between {\aox} and wind strength is driven by the relationship between the optical luminosity and wind strength. Our results are consistent with a radiation line-driven wind, whereby the ionising far-UV photons must not over-ionise the gas. The hard X-ray photons are few enough in number to have a negligible effect on the ionisation state of the material.

\end{abstract}




\section{Introduction}
\label{sec:intro}

Wide-angle quasar winds are a viable source for AGN feedback; however the driving mechanisms of such winds are unclear.
Accretion disc winds driven by radiation line driving is an established model. These winds can be launched at radii of 0.01\,pc \citep[e.g.,][]{murray_accretion_1995, proga_dynamics_2000}. Magnetic fields may be important at these small scales, while radiation pressure on dust is also a valid explanation with launching radii at torus scales \citep[e.g.,][]{scoville_stellar_1995, he_evidence_2022, ishibashi_are_2024}. See \citet{laha_ionized_2021} for a review.

Accretion-disc winds are often studied via the blueshift of the {\CIV}\,$\lambda1549$ emission line present in rest-frame UV quasar spectra (e.g., \citealt{gaskell_redshift_1982}; \citealt{sulentic_eigenvector_2000}; \citealt{richards_unification_2011}; but see \citealt{gaskell_line_2013, gaskell_case_2016} for an alternative view). 

\begin{figure}
    \centering
    \includegraphics[width=\linewidth]{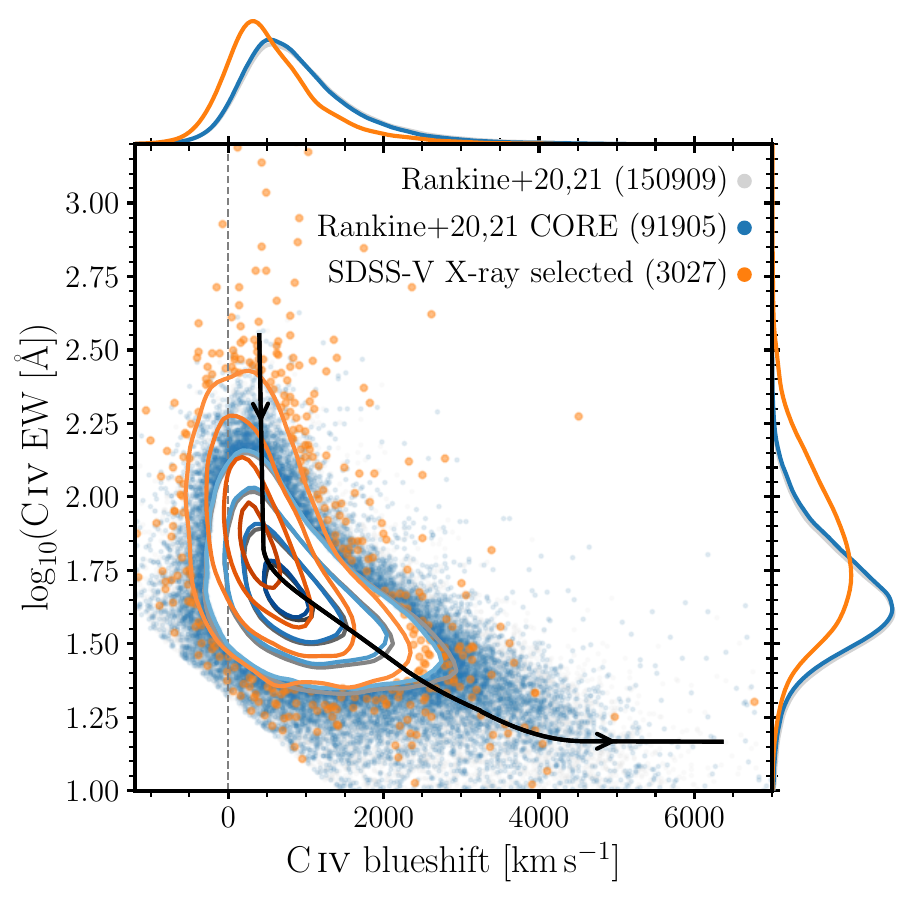}
    \caption{{\CIV} emission space for the SDSS-IV quasar sample presented in \citet{rankine_bal_2020, rankine_placing_2021} (grey) and the subset of this sample that was part of the CORE SDSS quasar targeting programmes (blue; see Section~\ref{sec:civ}). The contours encircle 11.8, 39.3, 67.5 and 86.4 per cent of the objects in each sample. The marginalised distributions have been renormalized. The {\CIV} distance reference line is the black curve, with the direction of increasing {\CIV} distance indicated. The optical selection that constitutes the CORE sample does not bias the SDSS-IV comparison sample towards a particular part of the parameter space. The X-ray selected SDSS-V sample (orange; discussed in Section~\ref{sec:civ}) is shifted towards lower blueshifts and larger {\CIV} EWs but spectra with significant blueshifts are still apparent in the sample.}
    \label{fig:CIV45}
\end{figure}

The {\CIV} emission space [i.e., the {\CIV} blueshift/asymmetry versus equivalent width (EW)] is populated by quasars with strong and symmetric {\CIV} emission, weak and symmetric, and weak but blueshifted emission \citep[e.g.,][]{richards_unification_2011}. \hlii{Objects lying towards the top-right of the space with strong and blueshifted {\CIV} emission are rarer, for example, the extremely red quasars discussed in} \citet{hamann_extremely_2017}, \hlii{and have been attributed to a larger covering factor of the {\CIV} emitting region than normal quasars}. The {\CIV} emission space is sometimes considered as part of the quasar (UV) main sequence. Figure~\ref{fig:CIV45} contains the {\CIV} emission space populated by the quasar sample studied by \citet[][hereafter \citetalias{rankine_bal_2020}]{rankine_bal_2020} and \citet{rankine_placing_2021} which was based on the SDSS DR14 quasar catalogue \citep{paris_sloan_2018}. This distribution is consistent with a line-driven accretion disc wind whereby quasars with harder spectral energy distributions (SEDs), therefore with stronger {\CIV} emission, are more likely to over-ionise the gas such that the cross-section for radiation line driving is decreased and no winds can be driven. Observationally, SED tracers such as the spectral slope between the 2500\,{\AA} monochromatic luminosity and 2\,keV X-ray luminosity \citep[{\aox};][]{tananbaum_x-ray_1979}, and the {\HeII} EW \citep{leighly_hubble_2004} have been shown to correlate with the location in {\CIV} emission space in a manner consistent with strong winds requiring softer SEDs (e.g., \citealt{baskin_average_2013, baskin_origins_2015, luo_x-ray_2015, ni_connecting_2018, ni_sensitive_2022, vietri_wissh_2018}; \citetalias{rankine_bal_2020}; \citealt{ timlin_correlations_2020, timliniii_what_2021,  rivera_exploring_2022}).

The non-linear correlation between {\HeII} and {\CIV} parameters \citepalias[e.g.,][]{rankine_bal_2020} led \citet{rivera_characterizing_2020} and \citet{rivera_exploring_2022} to define the {\CIVdist} (the black line in Fig.~\ref{fig:CIV45}). {\CIVdist} involves projecting each object on to the best-fitting polynomial through {\CIV} space and measuring the distance along the line. \hlii{As {\CIV} distance increases from zero, the EW decreases, and once EW $\lesssim1.75$, the blueshift starts to increase.} A convenience of the {\CIVdist} metric is that it reduces the {\CIV} emission space to one dimension. A consequence of its design is that {\HeII} EW and {\aox} are linearly correlated with {\CIVdist} \citep{rivera_exploring_2022}. The authors also surmise that {\CIVdist} is the most likely parameter to linearly correlate with the Eddington ratio, {\LLEdd} (where $L$ is the bolometric luminosity and $L_\text{Edd}$ the Eddington luminosity), but find little correlation with the X-ray spectral slope, $\Gamma$, despite $\Gamma$ itself being an {\LLEdd} indicator (e.g., \citealt{shemmer_hard_2008}; \citealt{brightman_statistical_2013}; \citealt{trefoloni_quasars_2024}; however, see \citealt{trakhtenbrot_bat_2017}; \citealt{wang_hot_2019}; \citealt{kamraj_x-ray_2022} for evidence suggesting only a weak $\Gamma$--{\LLEdd} relationship).

In addition to {\CIV} properties and {\LLEdd} being correlated \citep[e.g.,][]{bachev_average_2004, sulentic_c_2007} observations and simulations have revealed correlations between {\CIV} properties and mass accretion rate \citep[e.g.,][]{matthews_disc_2023}, black hole mass ({\BHM}; e.g., \citealt{temple_testing_2023}), bolometric luminosity \citep[the anti-correlation between EW and bolometric luminosity known as the Baldwin Effect;][]{baldwin_luminosity_1977}, the ionisation structure of the BLR \citep{temple_fe_2020, temple_high-ionization_2021}, hot nuclear dust contributions \citep{temple_exploring_2021,rivera_multiwavelength_2021, fawcett_fundamental_2022}, the ionised phase of the kiloparsec-scale interstellar medium \citep[e.g.,][]{coatman_kinematics_2019, temple_o_2024}, and the molecular phase \citep[tentatively;][]{molyneux_evidence_2025}.  
The framework of \citet{giustini_global_2019} suggests $L/L_\text{Edd}>0.25$ and $M_\text{BH}>10^8\,\text{M}_\odot$ must be satisfied for line-driven accretion disc winds to be launched.

The SED associated with any AGN disc wind model is generally thought to consist of three components: an optically thick accretion disc which emits thermally as a multi-component blackbody with a radially-dependent peak temperature; a hot Comptonised corona which dominates the X-ray emission above $\approx$1\,keV; and a warm Comptonising component which gives rise to the ``soft-excess'' of X-ray photons below 1\,keV \citep[e.g.,][]{kubota_physical_2018, mitchell_soux_2023}. The ratios of these components are dependent on {\BHM}, the accretion rate, and the spin of the black hole, and are evidenced by differences in the SED \citep[e.g.,][]{hagen_systematic_2024, waddell_erosita_2024}. A consequence of the effect on the SEDs is that samples containing optically selected blue quasars are likely to favour objects with bright accretion discs over objects with strong X-ray coronae, for instance.

Previous large spectroscopically-identified quasar samples have been predominantly optically selected, for example SDSS-I/II/III/IV \citep{york_sloan_2000, ross_sdss-iii_2012, myers_sdss-iv_2015} and DESI \citep{chaussidon_target_2023}.
SDSS-V \citep{kollmeier_sloan_2025} is targeting \textit{eROSITA}-selected sources \citep{Predehl_2021} as part of its Black Hole Mapper programme (Anderson et al. in preparation). 
In a recent paper, \citet{hiremath_x-ray_2025} examined the properties of the X-ray selected broad absorption line (BAL) quasars identified in the 19th data release of the SDSS \citep{collaboration_nineteenth_2025}, most of which were detected in eROSITA/DR1 \citep{Merloni_2024, Salvato_2025}. BALs are observed in $\sim$10 per cent of quasar spectra and are clear signs of quasar winds \citep[e.g.,][]{weymann_comparisons_1991}. They are most commonly identified by absorption in the {\CIV}\,$\lambda1549$ ion caused by outflowing gas intersecting our line-of-sight and there appears to be a connection between the BAL properties and the {\CIV} emission blueshift (e.g., \citetalias{rankine_bal_2020}; \citealt{matthews_disc_2023}), with a disc wind often invoked in both phenomena. \citet{hiremath_x-ray_2025} reported that the BAL quasars in their sample had X-ray properties similar to the non-BALs in the sample \citep[although note that other studies have found BAL quasars to be X-ray weaker than the non-BALs; e.g.,][]{brandt_2000, Luo2014, saccheo2023}. While their focus was the BAL quasar sample, they also revealed that the X-ray selected sample is biased towards lower {\CIV} emission blueshifts and larger rest equivalent widths (EWs) as expected by, e.g., \citet{timlin_correlations_2020}. 

In this paper, we investigate the {\CIV} \textit{emission} properties of the X-ray selected SDSS sample and to consider how the optical and X-ray properties correlate with the winds for this large sample with optical and X-ray information.

Section~\ref{sec:data} details the X-ray and optical data and sample selection as well as the improved redshift estimation and derivation of the optical and X-ray parameters used throughout this paper. Section~\ref{sec:civ} explores possible causes of the bias towards lower {\CIV} blueshifts found by \citet{hiremath_x-ray_2025}. Section~\ref{sec:wind_xray} presents our results regarding the link between the UV and optical properties and wind signatures probed in our sample and the lack of a connection to the X-ray properties. We discuss our results in the context of previous analyses in Section~\ref{sec:dis_lit}, possible explanations for the uncorrelated X-ray and wind properties in  Section~\ref{sec:dis_xray}, and the link to {\BHM} and {\LLEdd} in Section~\ref{sec:dis_bh}. We conclude with a summary of our results and future prospects in Section~\ref{sec:conc}. 
A $\Lambda$CDM cosmology with $h_0$ = 0.71, $\Omega_\textup{M}$ = 0.27, and $\Omega_{\Lambda}$ = 0.73 is adopted for determining quantities such as quasar luminosities. All EWs are rest-frame values.

\section{Data and sample selection}
\label{sec:data}

\hlii{We summarise our sample in Section~{\ref{sec:data_summary}}, but interested readers can refer to Section~{\ref{sec:opt}} for details regarding the SDSS-V spectra and derived optical/UV parameters including luminosities and emission line measurements; or Section~{\ref{sec:xray}} for details of the \textit{eROSITA} cross-match and X-ray spectral fitting.}

\subsection{Optical/UV}
\label{sec:opt}

To compile our X-ray selected quasar sample we start from the objects targeted by SDSS-V based on an X-ray detection in either \textit{eROSITA}, the Chandra Source Catalog \citep{evans_2024}, and the \textit{XMM-Newton} \citep{Webb_2020} or \textit{Swift}-XRT serendipitous catalogues \citep{delaney_extragalactic_2023}.  
Optical spectra were obtained using the BOSS multi-fibre spectrographs \citep{smee_2013} on the Apache Point Observatory 2.5\,m SDSS telescope \citep{Gunn_2006} and the 2.54\,m Irénée du Pont Telescope at Las Campanas Observatory \citep{bv73}. Our sample contains objects that were first observed up to an MJD of 60280. The objects with MJD $\leq60130$ are now public as part of the SDSS DR19 \citep{collaboration_nineteenth_2025}. The more recent data will be made public through DR20.
The spectra must also be classified as `QSO' by the SDSS BOSS pipeline, idlspec2d (v6\_2\_1; \citealt{bolton_2012}; Morrison et al., in preparation) and have ZWARNING=0 indicating no apparent issues with the redshift fitting were flagged by the pipeline. We consider only objects with redshifts $1.5\lesssim z \lesssim 3.5$ such that {\CIV} falls within the observed wavelength range, and that are within the \textit{eROSITA} footprint. With these selection criteria, our sample numbers 5510, prior to cross-matching with the deepest available \textit{eROSITA} X-ray catalogues as required for the X-ray spectral analysis (see Section~\ref{sec:xray_fit} below).

\subsubsection{Redshift corrections} 
\label{sec:zcorr}

The SDSS pipeline redshifts can be biased when the {\CIV} emission line is asymmetric and blueshifted. These systematic redshift differences can be significant compared to the blueshifts measured -- of order $1000$\,{\kms}.

\begin{figure*}
    \centering
    \includegraphics[width=0.8\linewidth]{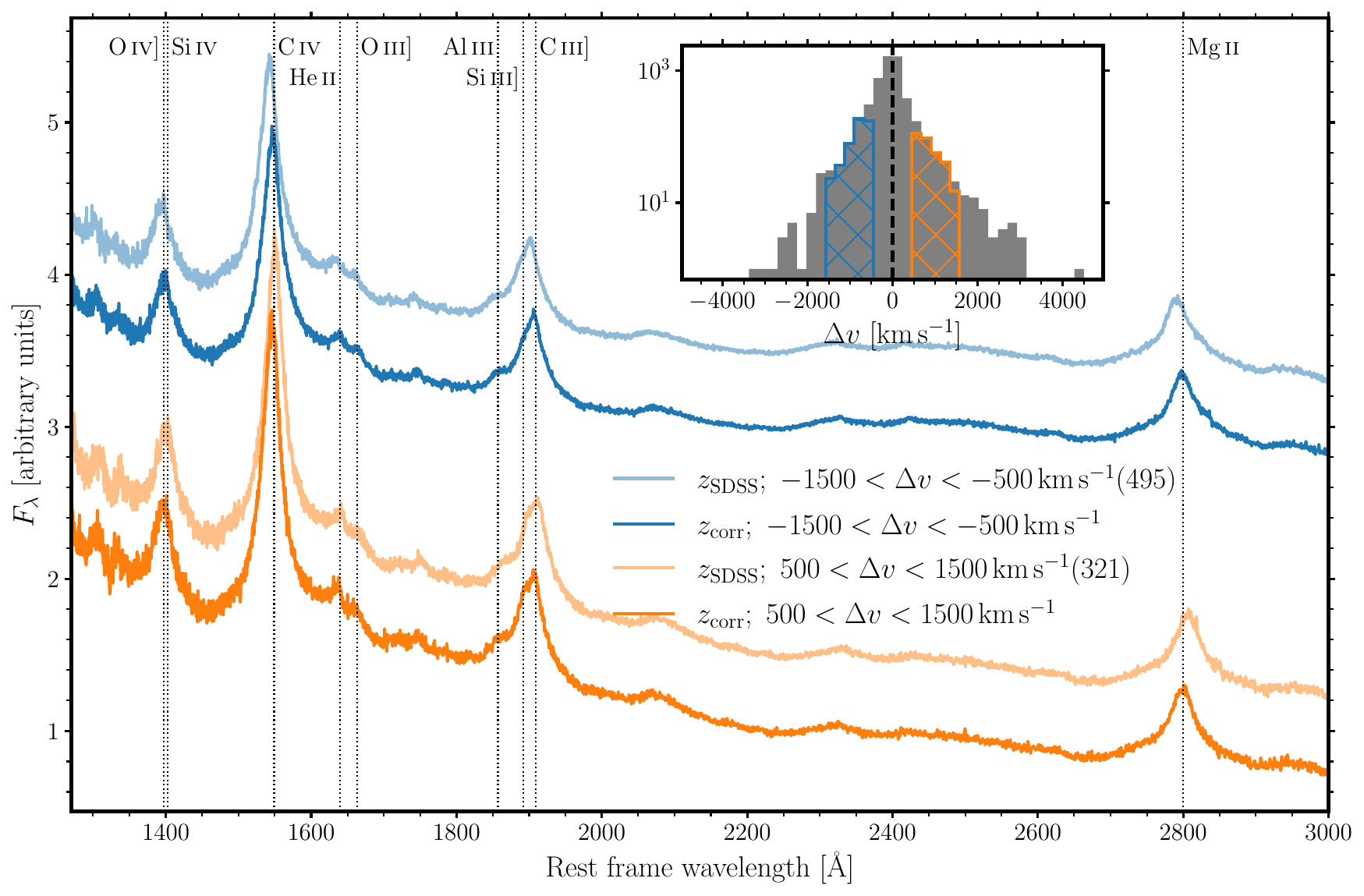}
    \caption{Rest-frame quasar spectrum composites generated using the SDSS pipeline redshifts (light blue and light orange) and updated redshifts from this work (dark blue and dark orange) for quasars with redshift differences $-1500 \leq \Delta v \leq -500$\,{\kms} (blue) and $500 \leq \Delta v \leq 1500$\,{\kms} (orange). The distribution of $\Delta v$ is presented in the inset figure. Most notably, the peak of the \ion{Mg}{ii}$\,\lambda$2800 emission line is shifted towards 2800\,{\AA} when using the corrected redshifts and the \ion{C}{iii}]\,$\lambda$1909 is in better alignment.}
    \label{fig:comp_z}
\end{figure*}

To improve the redshift estimates, we implement the cross-correlation algorithm outlined in section 4.2 of \citet{hewett_improved_2010} and updated in \citet{stepney_no_2023} with a set of 33 templates that span the full variety of emission-line profiles and cover the range 1265--3000\,{\AA}. The scheme is described in full across these two papers but we note that only the template region > $1675$\,{\AA} is used in the cross-correlation to avoid using {\CIV} in the fitting. The \ion{C}{iii}]\,$\lambda1909$ complex and \ion{Mg}{ii}\,$\lambda2800$ emission line dominate the redshift solution. Fig.~\ref{fig:comp_z} displays the improvement (dark blue and dark orange) upon the pipeline redshifts (light blue and light orange): the composite spectra show that we obtain, on average, better agreement between the vacuum wavelengths and peaks of the \ion{C}{iii}] and \ion{Mg}{ii} emission lines \citep[as expected; see][]{shen_sloan_2016}. The $\Delta v < 0$\footnote{$\Delta v = c(z_\text{corr} - z_\text{SDSS})/z_\text{corr}$.} regime (blue), represents improvements due to the increased number of templates covering a range of SEDs considered in the redshift estimation compared to the SDSS pipeline (four PCA-derived quasar templates). In the $\Delta v > 0$ regime (orange), the blueshifted {\CIV} emission leads the SDSS pipeline to estimate lower redshifts which underestimates the measured {\CIV} blueshifts. The inset figure shows the histogram of overall shifts, revealing a tail towards larger positive corrections due to large {\CIV} blueshifts.

\subsubsection{Optical/UV properties}
\label{sec:opt_prop}
We reconstruct the SDSS spectra following the ICA-based spectrum-reconstruction scheme described in \citetalias{rankine_bal_2020}. We make use of the same quasar components -- derived from a sample of SDSS-IV quasars -- and that \citet{hiremath_x-ray_2025} have shown recently are able to provide good \hlii{(primarily with reduced $\chi^2 \lesssim2$)} fits to the SDSS spectra of the X-ray selected SDSS-V quasar samples.

\hli{
As in {\citetalias{rankine_bal_2020}}, we measure various rest-frame UV properties from the reconstructions. For emission line measurements including {\CIV}, {\HeII} and {\ion{Mg}{ii}} (for calculating black hole masses), we first fit to the reconstruction localised powerlaw continua under the emission lines using the median flux in two wavelength regimes: 1445--1465 and 1700--1705\,{\AA} for {\CIV}\footnote{1610--1620 and 1700--1705\,{\AA} for {\HeII} \citep[identical to][]{baskin_average_2013}, and 2500--2650 and 2920--2975\,{\AA} for \ion{Mg}{ii}.}. The reconstructed emission line flux is then \hlii{directly integrated} over 1500--1600\,{\AA} for {\CIV}\footnote{1620--1650\,{\AA} for {\HeII}.}. The {\CIV} blueshift is calculated from the wavelength which bisects the cumulative total line flux, blueshifted with respect to the vacuum wavelength of the emission line, in this instance 1549.48\,{\AA} for the {\CIV} doublet.}

\hli{We estimate black hole masses ({\BHM}; in M$_\odot$) following the single-epoch estimator of} \citet{vestergaard_mass_2009} \hli{which uses the Full-Width at Half-Maximum (FWHM) of the {\ion{Mg}{ii}} emission line:}
\begin{equation}
    M_\text{BHM} = 10^{6.86}\left(\frac{\text{FWHM Mg\,{\textsc{ii}}}}{10^3\,\text{km\,s}^{-1}}\right)^2\left(\frac{L_{3000}}{10^{44}\, \text{erg\,s}^{-1}}\right)^{0.5}.
\end{equation}
\hli{{\ion{Mg}{ii}} is only present in the spectra with $z\lesssim2.57$, thus any investigations regarding the {\BHM} or Eddington ratio are limited to this redshift range (2730 quasars).}
\hlii{The 3000\,{\AA} monochromatic luminosity, $L_{3000}$, required above is measured from the spectrum reconstructions corrected for Galactic extinction with the \textsc{dustmaps} Python module} \citep{green_dustmaps_2018} \hlii{and the dust map of} \cite{Schlegel_maps_1998} \hlii{updated by} \citet{schlafly_measuring_2011} \hlii{in tandem with the \textsc{extinction} module} \citep{barbary_extinction_2016} \hlii{and the reddening curve of} \citet{fitzpatrick_correcting_1999}. \hlii{The 2500\,{\AA} monochromatic luminosity, $L_{2500}$, is generated in the same way.} \hlii{Bolometric luminosities are estimated from $L_{3000}$ assuming a bolometric correction of 5.15} \citep{shen_catalog_2011}.

{\HeII} is a weak line, thus uncertainties on the EW measurements will be dominated by the continuum placement, particularly in low S/N spectra. \hlii{We estimate the average uncertainty on the {\HeII} EW via Monte Carlo error estimation: for 500 randomly selected spectra, representing the distribution of S/N in our sample, we create 50 noisy realisations and repeat the reconstruction process and {\HeII} measurements. The median uncertainty on $\log_{10}({\HeII}\ \text{EW})$ is 0.03 dex. We also calculate the {\HeII} line luminosity, $L_{\HeII}$, from the reconstructions and calculate a median uncertainty on $\log_{10}L_{\HeII}$ of 0.03 dex. Fifty samples is sufficient to estimate the {\HeII} uncertainties with $\sim$10 per cent Monte Carlo uncertainty.}

We also calculate the Balnicity Index \citep{weymann_comparisons_1991} to determine which objects are broad absorption line (BAL) quasars. We find 207/2987 (7 per cent) BAL quasars in our sample, similar to the 6 per cent found in a similarly X-ray selected SDSS-V sample analysed in \citet{hiremath_x-ray_2025}. BAL quasars are found to be on average X-ray weak \citep[e.g.,][]{green_1995, laor_1997, brandt_2000, gallagher_exploratory_2006, gibson_catalog_2009, Luo2014, saccheo2023}; however, \citet{hiremath_x-ray_2025} reported that the SDSS-V X-ray selected BAL quasars have X-ray properties similar to their non-BAL counterparts. Based on this result and the fact that the reconstruction scheme allows an accurate recovery of the {\CIV} and {\HeII} emission line properties (see appendix B of \citetalias{rankine_bal_2020}), we include the BAL quasars in our analysis; excluding them does not qualitatively alter our conclusions.

\subsection{X-ray}
\label{sec:xray}

\subsubsection{Matching to \textit{eROSITA} catalogues}
\label{sec:xray_match}
To obtain information about the X-ray emission for the sources in the {\CIV} sample, we make use of the data collected by \textit{eROSITA} \citep{Predehl_2021}, aboard the spacecraft SRG \citep{Sunyaev_2021}. The \textit{eROSITA} dataset used here is limited to the German \textit{eROSITA} collaboration (eROSITA\_De), whose footprint covers half the sky, at Galactic longitudes $180\degree < l < 360\degree$. We make use of data from two separate \textit{eROSITA} surveys: the \textit{eROSITA} Equatorial Final Depth Survey \citep[eFEDS;][]{Brunner_2022} and the \textit{eROSITA} All-Sky Survey \citep[eRASS;][]{Merloni_2024}. eFEDS covers only a small region of the sky ($\sim$140~deg$^2$; $126\degree<\textrm{RA}<146\degree$ \& $-3\degree<\textrm{Dec.}<6\degree$), whereas the eRASS data cover the full eROSITA\_De footprint.

The counterparts of \textit{eROSITA} sources targeted by SDSS-V are defined in the two fields described in \citet{Salvato_2022,Salvato_2025}. Matching of the counterparts was done using NWAY, a Bayesian algorithm enhanced with a prior designed to identify X-ray emitters among data from Legacy Survey DR10 \citep{dey_overview_2019,Zenteno_2025}, CatWISE2020 \citep{CatWISE2020}, and Gaia DR3 \citep{GaiaDR3_summary}. To obtain the \textit{eROSITA} data for our SDSS-selected sample we make use of counterpart catalogues matched using the same method: the published eFEDS counterpart catalogue \citep{Salvato_2022} and a catalogue for eRASS4:5, available internally to the eROSITA\_De collaboration. The eRASS4:5 dataset contains the observations with the maximum depth available at the point the survey was halted. For further details of our matching procedure we refer the interested reader to Appendix~\ref{app:xray}. For our sample we identify X-ray counterparts for 4725/5510 (85 per cent) of the SDSS targets.

\subsubsection{X-ray spectral fitting}
\label{sec:xray_fit}

\begin{figure*}
    \centering
    \includegraphics[scale=0.45,trim=2cm 0.9cm 3cm 2.5cm, clip]{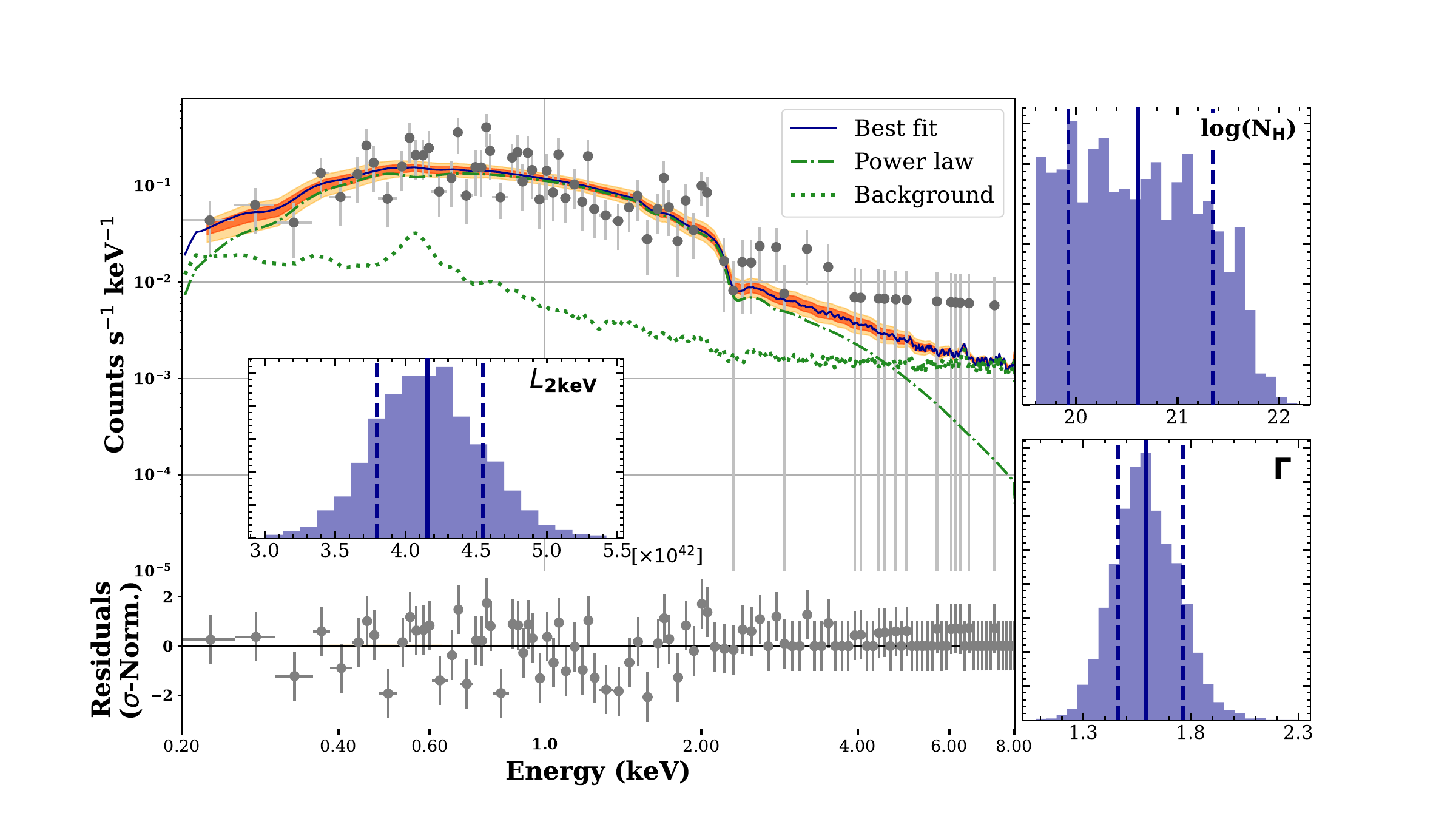}
    \caption{Example of our Bayesian X-ray modelling procedure. This is an eRASS:5 spectrum with a total of 253 photon counts and is for an AGN with SDSS fibre coordinates (hms) RA, Dec = 04:18:26.22, -03:18:41.9. In the upper left panel we show the spectral data in counts and the best-fitting model (\textit{dark blue, solid line}). The model consists of two main components: the source model (\texttt{TBabs*zTBabs*powerlaw}, \textit{green dotted}) and the model for the background (\textit{green dash-dot}). The confidence regions for the model are shown for 1$\sigma$ (\textit{orange}) and 99 per cent (\textit{yellow}) level confidence. The bottom left panel shows the error-normalised residuals of the model. We show the posterior constraints on the model parameters, specifically for the intrinsic absorption log(N$_\textrm{H}$ [$10^{22}$ $\textrm{cm}^{-2}]$) in the upper right panel, the power-law index $\Gamma$ in the lower right panel, and the derived 2~keV monochromatic luminosity ($L_{2\,\textrm{keV}}$ [$10^{42}$ erg~s$^{-1}$ keV$^{-1}$]) as inset in the upper left panel.
    Solid lines show the means of the distributions and dashed lines indicate the 1$\sigma$ intervals. The photon index and 2keV luminosity are typically better constrained than the absorption.}
    \label{fig:xray_fitting}
\end{figure*}

For the eFEDS data we make use of the published spectra\footnote{https://erosita.mpe.mpg.de/edr/eROSITAObservations/} and for the eRASS data we use the most recent stacked spectra (c030), available internally to the eROSITA collaboration. The spectral extraction is conducted using the standard eSASS pipeline and is described in detail in \citet{Brunner_2022} and \citet{Merloni_2024}. To fit the resulting source and background spectra, we follow the procedure set out in \citet{liu_erosita_2022}. We make use of a Bayesian fitting approach with XSpec \citep[][]{Arnaud_1996}, as implemented in the Bayesian X-ray Analysis package \citep[BXA;][]{Buchner_2014}. For the objects included in the eFEDS AGN catalog, we make use of the fitting results by \citet{liu_erosita_2022}, in the form of the full posterior distributions for the spectral parameters.
Where possible the eFEDS fits have been updated and re-run using the most up-to-date SDSS-V redshifts. For the eRASS-matched sources, we perform new X-ray fitting. We briefly describe the method here. 

The background is modelled phenomenologically, following the method by \citet{Simmonds_2018}. For each combined dataset (eFEDS-matched and eRASS-matched sources, respectively), a generic model of the X-ray background is created based on PCA analysis of the background spectra. Each individual background spectrum is subsequently fit with a model formed by a linear combination of the first six principle components, with the relative normalisations as free parameters. Once the PCA-model normalisations are fixed, Gaussian components are added to the model, which continues as long as the newly added component leads to an improvement to the fit based on the Akaike Information Criterion. Once the best background model is found, this is stored as an XSPEC model table, to be used in the fitting of the source spectrum. 

The source spectra are fit with a combination of the background model and an absorbed power-law, the XSpec model \texttt{TBabs*zTBabs*powerlaw}. For the Galactic extinction, \texttt{TBabs} \citep{Wilms_2000}, the total Hydrogen column density (N$_\textrm{H}$) is calculated following the procedure set out in \citet{Willingale_2013} \citep[who use \HI~measurements from][]{Bekthi_HI4PI_2016} and kept fixed in the fitting. For the intrinsic extinction, \texttt{zTBabs}, the redshift is fixed to the SDSS-V value. For the free parameters of the fit we adopt the same priors as \citet{liu_erosita_2022}: log-uniform priors for the intrinsic N$_\textrm{H}$ ($4 \cdot 10^{19}<$ N$_\textrm{H} [\textrm{cm}^{-2}] < 4 \cdot 10^{24}$) 
and the normalisation of the power-law,
a Gaussian prior with mean $\mu=2.0$ and standard deviation $\sigma=0.5$ for the power-law's spectral index ($\Gamma$), and a uniform prior for the normalisation of the otherwise fixed background model. 
The BXA fitting procedure calculates the posterior distributions for these four parameters. In Figure~\ref{fig:xray_fitting} we show an example of the results of our fitting procedure. The figure also shows the parameter constraints on spectral parameters we can draw based on the posterior distributions.

To determine the spectral slopes $\alpha_\text{ox}$ and $\alpha_\text{uvx}$ (see Section~\ref{sec:wind_xray}) we require the 2 keV monochromatic luminosity ($L_{2\,\textrm{keV}}$). We calculate this value in XSpec based on the 1.999--2.001 keV luminosity for the best-fitting model. In the case of the unmatched sources, we calculate upper limits of the flux. For the eFEDS objects we make use of the estimated 0.2--2.3 keV limits of $10^{-14}$ erg~cm$^{-2}$~s$^{-1}$ \citep{Brunner_2022}. In the case of the unmatched eRASS sources, we use the deepest available eRASS data and follow the procedure for calculating the upper limits defined in \citet{Tubin-Arenas_2024}. For eRASS we make use of observed counts and use the same energy band as for the eFEDS upper limits. Combining these results, we find a median upper limit of $4.1 \cdot 10^{-14}$ erg~cm$^{-2}$~s$^{-1}$ for the unmatched sources in our combined sample.

\subsubsection{Posterior constraints on the distribution of X-ray parameters for the sample}
\label{sec:xray_post}
The count rate for most \textit{eROSITA} spectra in our sample is relatively low,
which means that the posterior distributions found in our fitting are often quite wide. This makes finding meaningful constraints on the spectral parameters of individual sources difficult. However, we can extract information about the distribution of a parameter in the sample as a whole. The parameter of particular interest for the current study is $\Gamma$, indicating the effective powerlaw slope of the X-ray spectrum.
We assume that for each spectrum, $\Gamma$ 
is drawn from a single parent distribution. We can then infer properties of the parent distribution from the posterior distributions of the individual spectra, through Hierarchical Bayesian Modelling (HBM). We follow the procedure as used in \citet{liu_erosita_2022} and \citet{Baronchelli_2020}, implemented in the \texttt{PosteriorStacker} Python package\footnote{https://github.com/JohannesBuchner/PosteriorStacker}.

We assume that the parent distribution of the spectral index is Gaussian and aim to find the corresponding mean, $\mu$, and standard deviation, $\sigma$. For a spectral parameter $s$, the combined likelihood function for the sample is given by
\begin{equation}
\label{eq:likelihood}
\mathcal{L} = \prod_{i}\int P(s|D_i)N(s|\mu,\sigma)\mathrm{d}s,
\end{equation}
where $P(s|D_i)$ is the posterior distribution of $s$ for an individual spectrum $i$ with data $D_i$, $N(s|\mu,\sigma)$ is the Gaussian parent distribution, and the product is over all individual spectra. If we ensure a broad uniform prior on $s$, we can make use of importance sampling to perform the integration in Equation~\ref{eq:likelihood} numerically, through sampling of the individual posterior distributions. The resulting likelihood, depending only on $\mu$ and $\sigma$, is then sampled using the same multi-nesting approach as used for the Bayesian fitting. For a full description of the approach, we refer to the appendix of \citet{Baronchelli_2020}.

For our eFEDS sample, using the mean values and 1$\sigma$ standard deviations of the posterior distributions, we find the parent distribution has $\mu_\Gamma = 2.07\pm0.02$ and $\sigma_\Gamma = 0.05\pm0.01$. This is softer than the typical AGN spectral index of $\sim$1.7, but in agreement with the results found for the full eFEDS AGN sample \citep{liu_erosita_2022}. The over-representation of AGN with steeper spectral slopes is most likely due to an \textit{eROSITA} selection bias towards softer sources, as well as a degeneracy of N$_\textrm{H}$ (in \texttt{zTBabs}) and $\Gamma$. Combining the eFEDS and eRASS fitting results for our combined sample, we find $\mu_\Gamma = 2.10\pm 0.01$ and $\sigma_\Gamma = 0.09\pm 0.01$. We note that these values represent a steeper power law than set by the prior, with a narrower range, indicating the shape of the distribution is constrained by the data. Using HBM on subsets of our sample, binned in {\CIV} parameter space, we can provide an additional test for a possible correlation between AGN wind and X-ray parameters, in Section~\ref{sec:wind_xray}.

\subsection{Summary of sample selection}
\label{sec:data_summary}

Our final sample, which contains only objects with an average $S/N>2$ per SDSS spectrum pixel ($\Delta v = 69$\,\kms) and good spectrum reconstructions \hlii{(reduced $\chi^2\leq2$)}, numbers 3027. Of this sample, 213 are broad absorption line quasars and are retained in the sample (see Section~\ref{sec:opt_prop}). A total of 1235/3027 (41 per cent) objects were  targeted by SDSS-V due to being X-ray selected by \textit{eROSITA} \hlii{(via either the eRASS:1 or eRASS:3\footnote{\hlii{eRASS:3 will be published in summer 2026 by Ramos-Ceja et al. (in prep) using eSASS c030. The targeting catalogue for SDSS-V was compiled with eSASS c020. See Appendix~{\ref{app:xray}} for details.}} target catalogues, internal to the collaboration)} and are subsequently detected in eRASS:4 or eRASS:5 \citep[see][]{Merloni_2024, Salvato_2025}; 1792 (59 per cent) objects have detections in eFEDS \citep{Brunner_2022, Salvato_2022};
one additional source was targeted for being a Chandra source and has an \textit{eROSITA} detection; and four \textit{XMM-Newton}/\textit{Swift} sources have an \textit{eROSITA} detection.

\begin{figure}
    \centering
    \includegraphics[width=\linewidth]{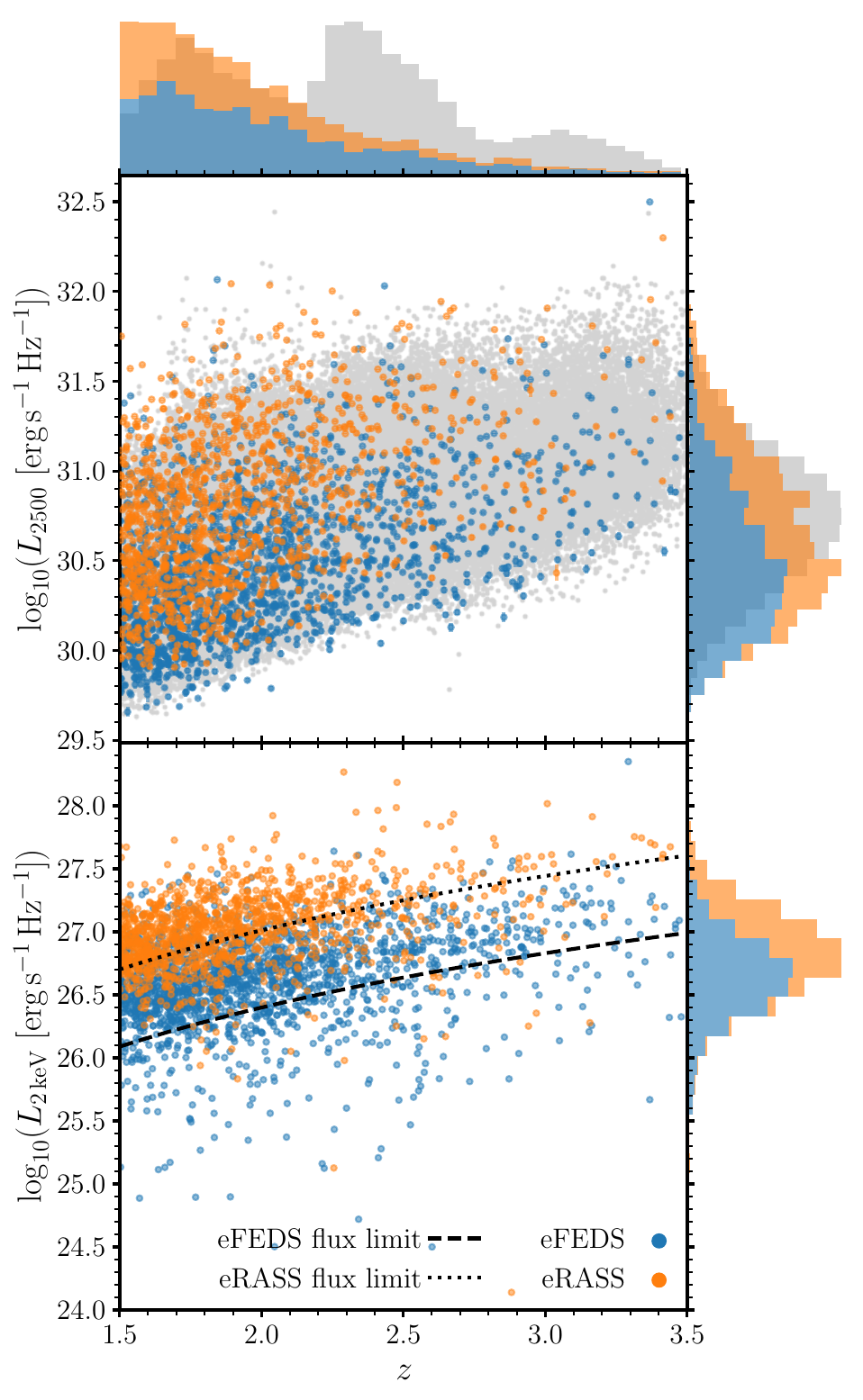}
    \caption{Optical (monochromatic luminosity at 2500\,{\AA} rest-frame wavelength, top) and X-ray (monochromatic luminosity at rest-frame 2\,keV, bottom) luminosities as a function of redshift for the X-ray selected SDSS-V quasar sample divided into eFEDS (blue) and eRASS (orange). 
    The optically selected SDSS-IV DR14 sample's $L_{2500}$ distribution is displayed in grey for comparison. The 1D distributions show the stacked SDSS-V samples. The eFEDS and eRASS nominal flux limits (calculated in Section~\ref{sec:xray_fit}), converted to 2\,keV luminosity limits assuming an X-ray spectrum with photon index $\Gamma=2$, are indicated as the dashed and dotted lines, respectively. 
    }
    \label{fig:lum}
\end{figure}

Figure~\ref{fig:lum} contains the 2500\,{\AA} (top panel) and 2\,keV (bottom panel) monochromatic luminosities as a function of redshift for our sample. For comparison, the \citetalias{rankine_bal_2020} optically selected sample is also presented in the top panel. The redshift distribution of the \citetalias{rankine_bal_2020} sample is complex and contains proportionally more objects at $z>2.2$ than our SDSS-V sample due to the various optical-colour selection criteria used to select quasar targets in prior SDSS surveys. Our SDSS-V sample is based purely on an X-ray selection (plus an optical magnitude limit). The samples cover the same $1.5\le z \le 3.5$ range. Our sample is weighted slightly towards lower optical luminosities, with a median $L_{2500}$ of $10^{30.58}$\,erg\,s$^{-1}$\,Hz$^{-1}$ driven by the deep X-ray eFEDS subsample, compared to $10^{30.74}$\,erg\,s$^{-1}$\,Hz$^{-1}$ for the \citetalias{rankine_bal_2020} sample. The average X-ray luminosity of the eFEDS sources is lower because this survey region probes deeper than the eRASS4:5 data (see the eFEDS and eRASS flux limits indicated on Figure~\ref{fig:lum}). 

\begin{figure}
    \centering
    \includegraphics[width=\linewidth]{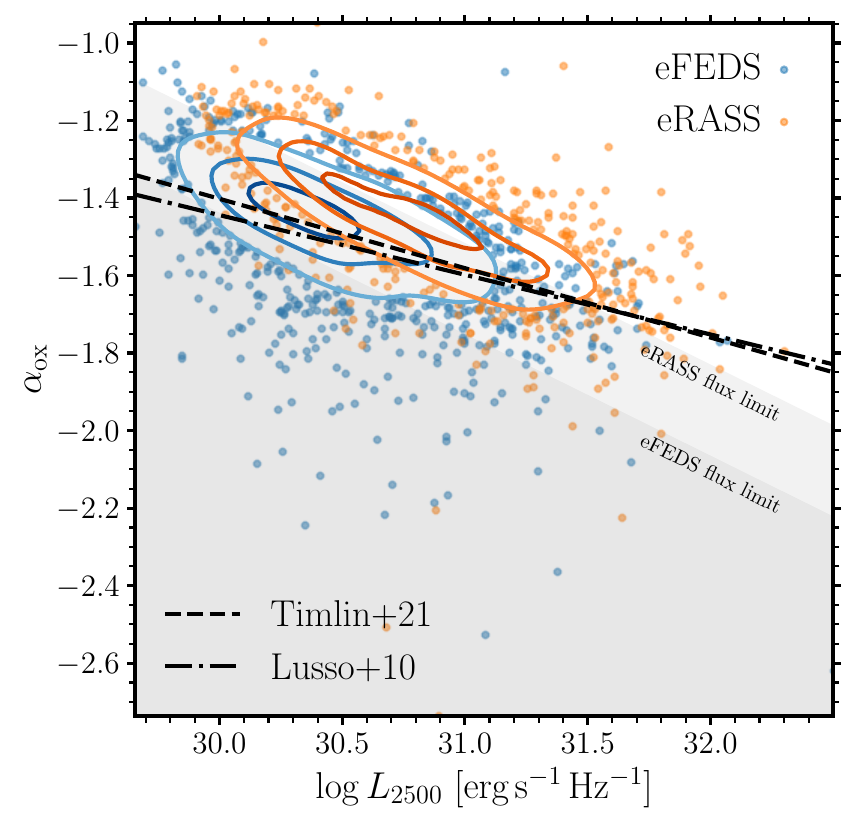}
    \caption{{\aox} as a function of $L_{2500}$ for the same subsamples as Fig.~\ref{fig:lum}. The grey regions demarcate the eFEDS and eRASS flux limits for $\Gamma=2$, converted to average {\aox} limits for our samples. The slopes and normalisations differ between the eFEDS and eRASS samples due to the different X-ray flux limits and area coverage, thus volume. The relations from \citet{timliniii_what_2021} and \citet{lusso_x-ray_2010} are indicated by the black dashed and dot-dashed lines, respectively.
    }
    \label{fig:aox_L2500}
\end{figure}

With access to the X-ray and optical information we can calculate the spectral slope between the monochromatic luminosities at rest frame 2\,keV and 2500\,{\AA}, {\aox} \citep{tananbaum_x-ray_1979}:
\begin{equation}
    \alpha_\text{ox} = \frac{\log_{10}\left(L_{2\,\text{keV}} / L_{2500\,\text{\AA}}\right)}{\log_{10}\left(\nu_{2\,\text{keV}} / \nu_{2500\,\text{\AA}}\right)},
    \label{eq:aox}
\end{equation}
where $\nu_{2\text{keV}}$ and $\nu_{2500\text{\AA}}$ are the frequencies. Fig.~\ref{fig:aox_L2500} shows the {\aox}--$L_{2500}$ parameter space for our sample, alongside the relation from \citet{timliniii_what_2021} which is given by $\alpha_\text{ox} = -0.179 \log L_{2500} + 3.968$ and derived from an optically selected sample with deep \textit{Chandra} observations. \hli{We also include the relation based on the X-ray selected sample of} \citet{lusso_quasars_2020}: \hli{$\alpha_\text{ox} = -0.154 \log L_{2500} + 3.176$. Our sample is less complete at low optical luminosities due to the sensitivity limit of the eRASS and eFEDS samples (note the intersection between the flux limits and the relations). }

\section[The effect of X-ray selection on the distribution in CIV space]{The effect of X-ray selection on the distribution in C\,{\sevensize IV} space}
\label{sec:civ}

We compare the distribution of the SDSS-V X-ray selected sample in {\CIV} emission space to that of the SDSS-IV optically selected sample by selecting the CORE objects from the DR14 quasar catalogue \citep{paris_sloan_2018} that were analysed in \citetalias{rankine_bal_2020} and \citetalias{rankine_placing_2021} (see Fig.~\ref{fig:CIV45}). This CORE sample (blue in Fig.~\ref{fig:CIV45}) is essentially identical to the full sample presented in \citetalias{rankine_bal_2020} and supplemented by \citetalias{rankine_placing_2021} (grey in Fig.~\ref{fig:CIV45}).
The X-ray selected sample covers the full {\CIV} parameter space; however, the contours and marginalised distributions reveal a shift towards higher EW and lower blueshifts compared to the optically selected quasar sample \citep[as shown by][for a similar sample]{hiremath_x-ray_2025}.
To investigate if the lack of high-blueshift objects is due to our much smaller sample size than the $\sim$92\,000-large optically selected sample, we perform bootstrap sampling by randomly selecting 3027 objects from the SDSS-IV sample and repeat 100 times. The bootstrapped samples still contain objects with large blueshifts. Separately, we perform a two-sample 2-D Kolmogorov–Smirnov test on the joint blueshift and EW distributions using the public code \textsc{ndtest}\footnote{Written by Zhaozhou Li, \url{https://github.com/syrte/ndtest}} \citep[algorithms based on][]{peacock_two-dimensional_1983, fasano_multidimensional_1987, Press2007}. The associated $p$-value\footnote{We reject the null hypothesis that the two samples were drawn from the same distribution when $p$-value~$<0.05$.}~$\ll3\times10^{-7}$ (5$\sigma$ equivalent) and the bootstrap sampling test both suggest that the optically selected and X-ray selected samples are not drawn from the same underlying population, or are drawn in different ways with different biases. Indeed, \citet{menzel_spectroscopic_2016} and \citet{Salvato_2025} report that $\sim50$ per cent of the AGN in their \textit{XMM} and \textit{eROSITA}-selected samples are unique to their X-ray selections.

\begin{figure}
    \centering
    \includegraphics[width=\linewidth]{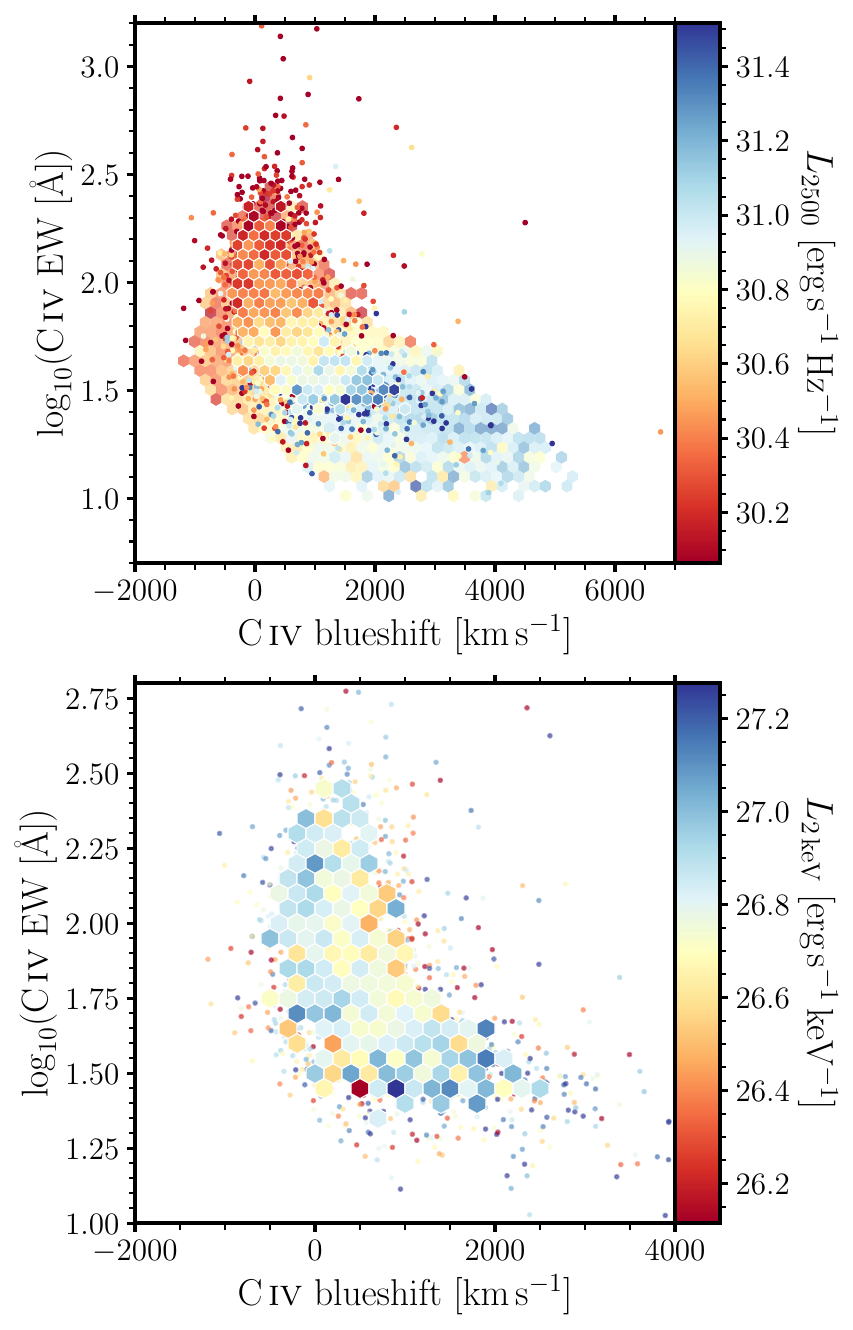}
    \caption{{\CIV} emission space colour-coded by the median optical luminosity per hexagon, $L_{2500}$ (top) and 2\,keV X-ray luminosity, $L_{2\,\text{keV}}$ (bottom). Note the differing axis limits. All hexagons contain at least 5 objects. In the top panel, the SDSS-IV sample is presented underneath the SDSS-V sample which is displayed as white-edged hexagons and dots. The optical luminosity is correlated with the {\CIV} emission space and the values are consistent across the optically and X-ray selected populations when their location in {\CIV} emission space is accounted for. The X-ray luminosity, however, does not appear to be correlated with {\CIV} properties.}
    \label{fig:CIV_L}
\end{figure}

In the remainder of this section, we consider what is leading to these different observed populations between our X-ray selected sample and the optically selected sample. We first discuss and investigate the effect of the observed redshift, optical luminosity, and Eddington ratio distributions in turn:
\begin{enumerate}
    
\item The redshift distributions of the two samples differ due to the complex targeting strategies of previous SDSS surveys (see Fig.~\ref{fig:lum}). 
Many investigations of high-redshift quasars show a bias towards larger {\CIV} blueshifts \citep[e.g.,][]{schindler_x-shooteralma_2020, belladitta_discovery_2025}; however \citet{stepney_no_2023} suggest that this behaviour is due to studies probing larger Eddington ratios at higher redshift instead of true redshift evolution of the wind properties. 

\item Although the X-ray selection leads to a different occupation of the {\CIV} emission space, the {\CIV} parameters show no correlation with {\Lx} (see bottom panel of Fig.~\ref{fig:CIV_L}). However, there is a clear gradient in $L_{2500}$ (top panel) across this space \citep[consistent with the well-known Baldwin Effect: the observed anti-correlation between optical luminosity and EW;][]{baldwin_luminosity_1977}. Both the optically and X-ray selected populations show similar correlations with the location in {\CIV} space, predicting the same median optical luminosity for a given set of {\CIV} properties.
In addition to being bright enough at X-ray wavelengths to be detected by \textit{eROSITA}, the sources in our sample must also be optically bright such that they meet the optical magnitude requirement for SDSS-V spectroscopy ($r<22.5$)\footnote{The SDSS-V spectroscopic magnitude limit is fainter than for previous SDSS surveys. Magnitude limits varied from $i<19.1$ in SDSS-I/II (\citealt{richards_spectroscopic_2002}) to $r<22$ by SDSS-IV (\citealt{lyke_sloan_2020}).}. Considering the distribution in optical luminosity in Fig.~\ref{fig:lum}, the X-ray selected population is on average, optically fainter than the optically selected population; however, this factor is only $\sim$0.3\,dex in the median. 

\item The \textit{eROSITA} selection favours softer X-ray sources, which can be associated with higher Eddington ratios (see Section~\ref{sec:intro}). However, our X-ray selected sample is statistically but mildly biased to lower Eddington ratios, with a median and standard deviation of the distribution of $\log\lambda_\text{Edd}\simeq-0.89$ and 0.35, respectively, while the optically selected sample has $\log\lambda_\text{Edd}\simeq-0.82$ and a width of 0.28. This result is perhaps due to the X-ray selection identifying AGN that are comparatively fainter relative to their host galaxies, and thus typically lower Eddington ratio objects given galaxy--BH scaling relations \citep{kormendy_coevolution_2013}. We would expect, based on \citetalias{rankine_bal_2020}, that lower Eddington ratios would bias our sample towards lower {\CIV} blueshifts. 
\end{enumerate}

We first check for redshift evolution of the {\CIV} properties within our sample by splitting the sample into two redshift bins at $z=2.25$ and applying a threshold of $L_{2500}>10^{30.5}$\,erg\,s$^{1-}$\,Hz$^{-1}$ to reduce luminosity-dependent effects. The 2-D KS test for the {\CIV} space distributions produces a $p$-value~$\approx0.04$ which is close to the threshold at which the null hypothesis -- that the two samples are drawn from the same underlying distribution -- cannot be rejected. We therefore conclude that true redshift evolution is not influencing the {\CIV} properties of our sample.

We investigate if any of the differing redshift, $L_{2500}$, and Eddington ratio distributions are the cause of the different occupation in {\CIV} space by creating optically selected samples matched in each of the properties to our X-ray selected sample. We select the 10 nearest SDSS-IV objects to each X-ray selected object in $z$, ($z$, $L_{2500}$), and ($z$, $\lambda_\text{Edd}$) and repeat the 2-D KS test for the {\CIV} space distribution. In all three cases, the $p$-value~$\ll3\times10^{-7}$, thus neither the redshift, optical magnitude limit, or the Eddington ratio can explain the observed {\CIV} distributions. The {\CIV} space for the ($z$, $L_{2500}$)-matched test is presented in Appendix~\ref{app:zLmatch}.

The lack of correlation between the X-ray luminosity and {\CIV} properties and the lack of high-blueshift objects could be a result of the sensitivity of \textit{eROSITA}. This sensitivity differs between eFEDS and eRASS, thus the bias may differ as well. We investigate this hypothesis by assigning mock X-ray fluxes to the optically selected sample based on the \citet{timliniii_what_2021} $L_{2500}$--{\aox} relation and applying the \textit{eROSITA} flux limits as thresholds for detection. See Appendix~\ref{app:mock_4} for details. The objects that fall below the flux limit are primarily at low {\CIV} blueshifts, thus the \textit{eROSITA} sensitivity alone is not sufficient to explain the lack of high blueshift objects in our sample. 

While there are clear and significant discrepancies between the X-ray and optically selected samples' occupation of the {\CIV} emission space, the tests we have performed indicate that this is \emph{not} purely a selection bias and thus that differences in intrinsic properties are responsible for the lack of the highest blueshift objects in our sample. Section~\ref{sec:wind_xray} explores the connection between the optical, UV, and X-ray emission, and the strength of the quasar winds as probed by the {\CIV} emission line in an effort to gain further insight to the X-ray selected sample's behaviour in {\CIV} space as well as the relationship between these three luminosities and winds.

\section[CIV winds as a function of X-ray, UV and optical properties]{C\,{\sevensize IV} winds as a function of X-ray, UV and optical properties}
\label{sec:wind_xray}

\begin{figure}
    \centering
    \includegraphics[width=\linewidth]{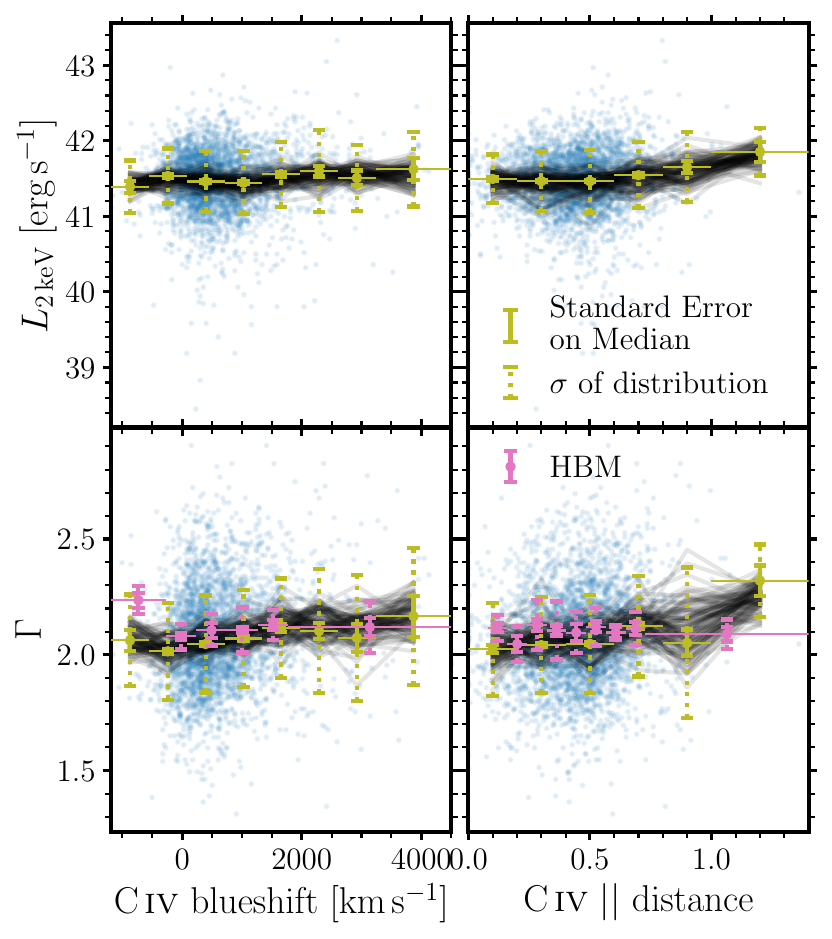}
    \caption{$L_\text{2\,keV}$ top, and X-ray spectral slope, $\Gamma$ (bottom), as functions of {\CIV} blueshift (left column) and {\CIVdist} (right column). Green circles and vertical error bars are the median, standard error (solid error bars), and standard deviation (dotted error bars) of the X-ray parameter in {\CIV} blueshift/distance bins. Horizontal error bars show the width of each bin. The pink points in the $\Gamma$ panel are the results of the Hierarchical Bayesian modelling (see text for details).
    For each panel, 100 bootstrapped curves are created in grey by randomly sampling N objects per bin, recording the mean, and repeating this procedure 100 times. N is the minimum number of objects in a bin per {\CIV} metric. None of the properties calculated from the X-ray spectra are strongly correlated with {\CIV} emission properties (see also Table~\ref{tab:spear_pears}).
    }
    \label{fig:CIV_xray}
\end{figure}

In this section we investigate the interplay between the optical, UV and X-ray emission and their effect on winds as traced by the {\CIV} emission line.
Figure~\ref{fig:CIV_xray} displays the parameters derived from the \textit{eROSITA} spectra, namely $L_\text{2\,keV}$, and X-ray spectral slope, $\Gamma$, as functions of {\CIV} blueshift and {\CIVdist} (see Section~\ref{sec:intro}).
We calculate {\CIVdist} for our X-ray selected sample using the public code \textsc{CIVdistance}\footnote{\url{https://doi.org/10.17918/CIVdistance}} \citep{CIVdist}. Further details can be found in \citet{rivera_characterizing_2020, rivera_exploring_2022, richards_probing_2021}.
None of the X-ray parameters correlate strongly with either {\CIV} blueshift or {\CIVdist}. The green points in the figure are the median values in bins of {\CIV} metric. 
The lack of a correlation between {\CIV} blueshift and $\Gamma$ is in agreement with the results of \citet{zappacosta_wissh_2020} who studied the sample of $z\sim2-4$ WISE/SDSS-selected Hyper-luminous quasars (WISSH). \hliii{However, they found a correlation between the 2--10\,keV X-ray luminosity and {\CIV} blueshift. In contrast, at $z>6$,} \citet{tortosa_hyperion_2024} \hliii{do observe a correlation in the HYPERION quasar sample between {\CIV} blueshift and $\Gamma$, but no significant relation with the 2--10\,keV luminosity.} We note that both papers define the blueshift as the shift of the peak of the line. As observed in Fig.~\ref{fig:aox_L2500}, the eRASS sub-sample in particular is biased against low X-ray luminosity sources. Appendix~\ref{app:mock_4} illustrates that, based on the relations in the literature between {\Lx} and $L_{2500}$, there may be a mild relationship between {\CIV} blueshifts/distances and {\Lx} that is lost when the eROSITA flux limits are applied. \hlii{Additionally, the lack of high-blueshift objects in our sample may reduce the observed trends if the X-ray properties only change at {\CIV} blueshifts $\gtrsim3000$\,{\kms}} \citep{shlentsova_x-ray_2026}.

The Spearman and Pearson coefficients listed in Table~\ref{tab:spear_pears} indicate that there are no clear correlations between the X-ray parameters and the wind strength as traced by the {\CIV} emission line. However, as noted in Section~\ref{sec:xray_post}, using the parameters derived from fits to individual X-ray spectra can be problematic when dealing with relatively low-count observations. For this reason we further test the correlation using Hierarchical Bayesian Modelling, under the same assumptions as in Section~\ref{sec:xray_post}. We bin the \textit{eROSITA} sources based on their {\CIVdist} and {\CIV}~blueshift, creating subsamples of at least 200 spectra each. For these subsamples we infer the mean and standard deviation of the parent distributions of the spectral parameter, $\Gamma$ (pink squares and error bars in Fig.~\ref{fig:CIV_xray}). Based on this analysis there is no significant trend in the spectral parameters with either of the {\CIV} measures.

\begin{figure}
    \centering
    \includegraphics[width=\linewidth]{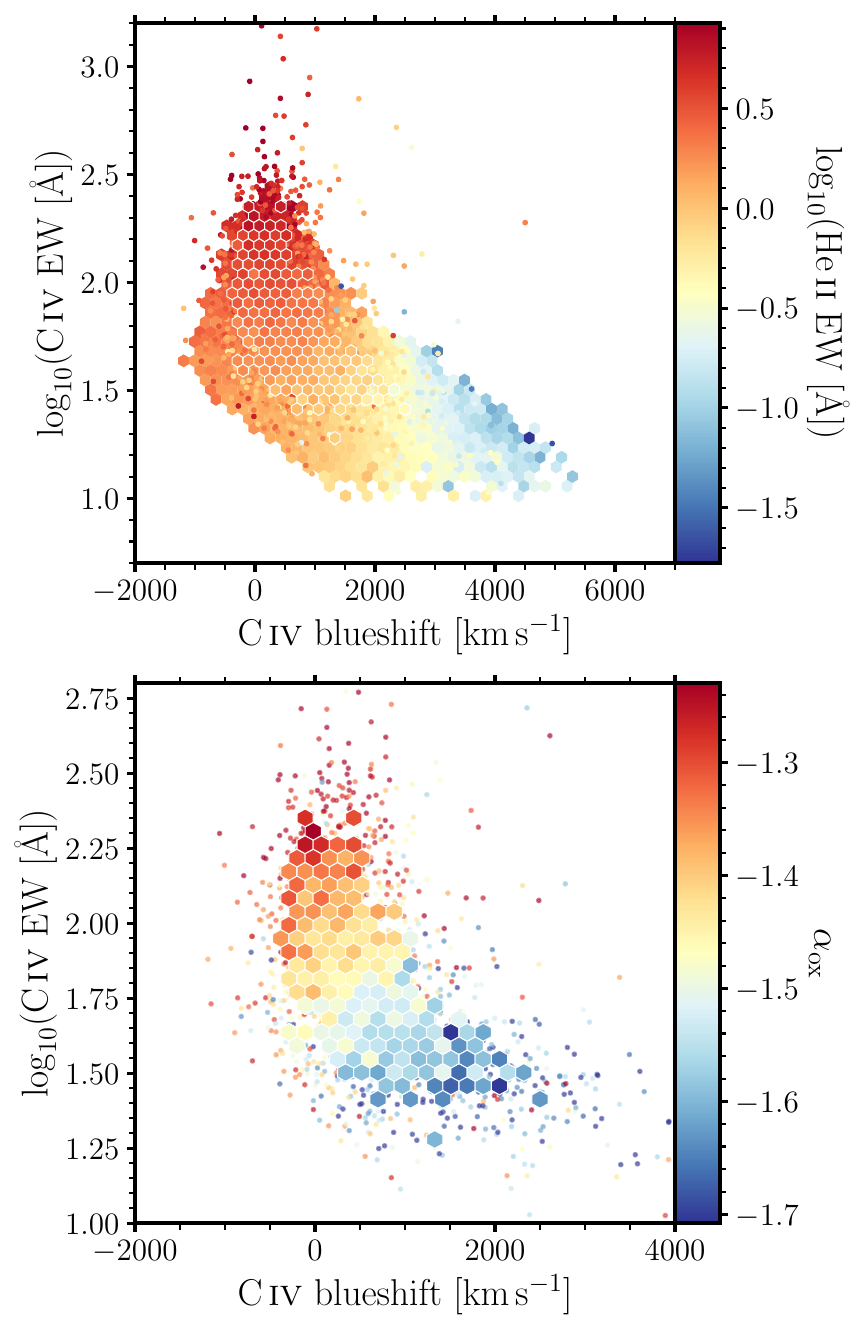}
    \caption{Same as for Fig.~\ref{fig:CIV_L} but colour-coded by the median {\HeII} EW (top) and optical-to-X-ray spectral slope, {\aox} (bottom). The {\HeII} properties of the X-ray selected population (hexagons outlined in white and circles) are consistent with those of the optically selected population when considering their location in {\CIV} emission space. The parameter {\aox} is strongly correlated with the {\CIV} emission properties with quasars with larger blueshifts having more negative {\aox} values (less X-ray emission relative to optical emission). This strong correlation is driven by changes in the optical luminosity (see Fig.~\ref{fig:CIV_L}).}
    \label{fig:CIV_HeII}
\end{figure}

{\HeII} is often used as an indicator of the number of ionising photons above 54\,eV. Specifically, high {\HeII} EW is indicative of a stronger soft X-ray spectrum \citep[][although in the sample discussed here we do not observe a correlation between {\HeII} EW and $\Gamma$]{leighly_hubble_2004} and has been shown to be correlated with {\CIV} blueshift (\citealt{baskin_average_2013, baskin_origins_2015}; \citetalias{rankine_bal_2020}). Figure~\ref{fig:CIV_HeII} presents the {\CIV} emission space but colour-coded by {\HeII} EW where the {\HeII} properties are as expected based on their location in {\CIV} space when compared to the optically selected sample.

\citet{timlin_correlations_2020, timliniii_what_2021} and \citet{rivera_exploring_2022} have shown that {\HeII} EW and {\CIV} blueshift are positively and negatively correlated with {\aox}, respectively. Figure~\ref{fig:CIV_HeII} confirms this result for our X-ray selected sample. As revealed in Fig.~\ref{fig:CIV_L}, it is variations in the optical luminosity that are driving the {\aox} trend with {\CIV} space.

It is not possible to measure the far-UV emission directly but we can use the {\HeII} line luminosity, $L_\text{He\,\textsc{ii}}$ (Section~\ref{sec:opt_prop}), as a proxy for the 54\,eV far-UV continuum luminosity \citep{mathews_what_1987}. As a recombination line, {\HeII} can essentially be used a counter of photons with energy $>54.4$\,eV. We estimate the 54\,eV continuum luminosity ($L_\text{54\,eV}$) by considering that the $L_\text{He\,\textsc{ii}}$ is proportional to the ionizing photon production rate above 54.4\,eV:
\begin{equation}
    Q(>54.4\,\text{eV}) = \frac{L_\text{He\,\textsc{ii}}}{h\nu_{1640} \frac{\Omega}{4\pi}},
\end{equation}
where $h\nu_{1640}$ is the energy of one {\HeII}\,$\lambda1640$ photon and $\Omega/4\pi$ is the covering fraction of the broad line region which we assume to be a constant 10 per cent. Using these assumptions one can estimate the continuum luminosity from
\begin{equation}
    Q(>54.4\,\text{eV}) = \int_{\nu_{54}}^{\infty} \frac{L_\nu}{h\nu}\,\text{d}\nu,
\end{equation}
which leads to a dependence on {\HeII} line luminosity of
\begin{equation}
    L_\text{54\,eV} = \frac{L_\text{He\,\textsc{ii}}}{\nu_{1640} \frac{\Omega}{4\pi}\alpha},
\end{equation}
where we have assumed a power-law SED: $L_\nu\propto\nu^{-\alpha}$ with $\alpha=3$ \citep[as per][]{mathews_what_1987}.
This estimation also requires the {\HeII} continuum to be optically thick.

\begin{figure}
    \centering
    \includegraphics[width=\linewidth]{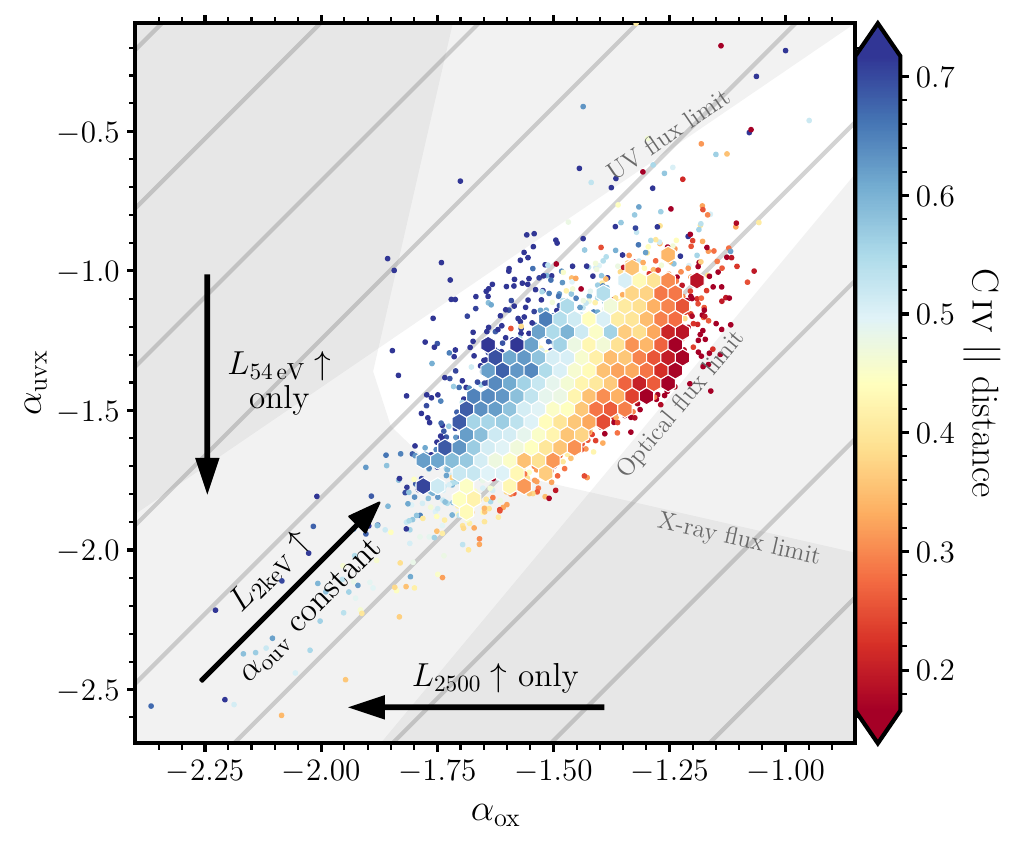}
    \caption{{\auvx} against {\aox} colour-coded by {\CIVdist}. Large arrows show how the optical, UV, and X-ray luminosity change across the axes. Increasing/decreasing the X-ray luminosity while holding {\aouv} constant produces movement along the grey lines towards top-right/bottom-left. The grey regions indicate the impact of the flux limits in this space (see text for details). The axes are scaled such that the same {\Lx} range is visible. Changing the X-ray luminosity alone is not sufficient to strongly affect the {\CIV} distance; instead, one must increase the optical luminosity and, with lesser significance, decrease the UV luminosity.}
    \label{fig:aox_auvx}
\end{figure}

With $L_{2500}$, $L_\text{54\,eV}$, and $L_\text{2\,keV}$ we can investigate how changes in these quantities affect the strength of accretion disc winds as probed by the {\CIV} emission space. In a similar manner to the {\aox} parameter (equation~\ref{eq:aox}), we define {\aouv} and {\auvx} as the spectral slopes between the optical and far-UV and the far-UV and X-ray, respectively:
\begin{equation}
    \alpha_\text{ouv} = \frac{\log_{10}\left(L_{54\,\text{eV}} / L_{2500\,\text{\AA}}\right)}{\log_{10}\left(\nu_{54\,\text{eV}} / \nu_{2500\,\text{\AA}}\right)},
\end{equation}
and
\begin{equation}
    \alpha_\text{uvx} = \frac{\log_{10}\left(L_{2\,\text{keV}} / L_{54\,\text{eV}}\right)}{\log_{10}\left(\nu_{2\,\text{keV}} / \nu_{54\,\text{eV}}\right)}.
\end{equation}
Figure~\ref{fig:aox_auvx} presents the {\auvx} against the classic {\aox}. Movement along the grey lines is achieved by changing $L_{2\,\text{keV}}$ while holding {\aouv} constant. The grey shaded regions mark the areas that are typically not accessible by our sample. \hlii{The X-ray limit is based on the average flux limit of the eFEDS and eRASS samples (see Section~{\ref{sec:xray_fit}}): we bin our sample in {\aouv} (demarcated by the grey diagonal lines in Fig.~{\ref{fig:aox_auvx}}) and calculate the average {\aox} and {\auvx} assuming the \textit{eROSITA} flux limit and the optical and UV luminosities in each {\aouv} bin. This approach leads to the curve in the X-ray limit. The optical flux limit is based on the $r$-band limit of 22.5, converted to luminosity at the redshifts of our sample and averaged in bins of {\auvx}. In a similar manner we estimate the UV flux limit using the minimum {\HeII} line flux we measure in the SDSS-IV sample, which is converted to a luminosity limit at each redshift in our sample and averaged in bins of {\aox}. The limits in {\aox}--{\auvx} space should only be considered qualitatively given the many layers of abstraction from the quoted flux/magnitude limits.}

\begin{figure}
    \centering
    \includegraphics[width=\linewidth]{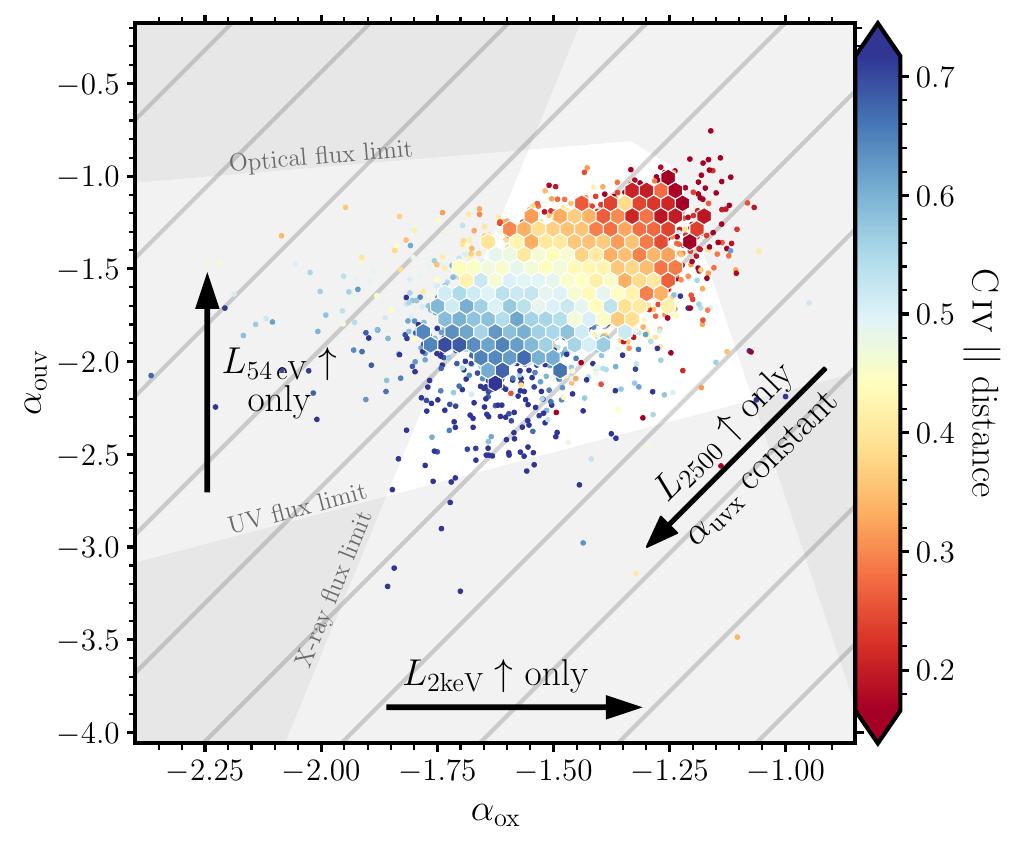}
    \caption{Similar to Fig.~\ref{fig:aox_auvx} but for {\aouv} against {\aox}. As shown in Fig.~\ref{fig:aox_auvx}, changing the X-ray luminosity alone does not substantially impact the {\CIVdist}. The ratio of optical-to-UV luminosity is important for affecting the most significant changes in {\CIVdist}, driven mainly by the increasing optical luminosity.}
    \label{fig:aox_aouv}
\end{figure}

The strong correlation between {\auvx} and {\aox} (Spearman rank coefficient of 0.66 and $p$-value~$\ll3\times10^{-7}$) is a result of the 2\,keV X-ray luminosity appearing in both axes. However, the correlation between these two parameters is stronger compared to the correlation between {\aouv} and {\aox} where $L_{2500}$ appears in both parameters (Spearman rank coefficient = 0.48, $p$-value~$\ll3\times10^{-7}$; see Fig.~{\ref{fig:aox_aouv}}).
As expected, the {\CIVdist} is correlated with the traditional {\aox} in Fig.~{\ref{fig:aox_auvx}}, with stronger winds in objects with the most negative {\aox} values. We also observe stronger winds when the UV luminosity is weaker relative to the X-ray (moving upwards in the figure). However, the steepest gradient in change of the {\CIVdist} is achieved by moving nearly perpendicular to the lines of changing X-ray luminosity. In other words, the X-ray luminosity does not need to change to increase the wind strength but instead the UV luminosity must decrease and the optical luminosity increase both relative to the X-ray luminosity. \hlii{In fact, the optical luminosity has a key role in driving the strongest variations in the {\CIV} distance and this is shown more clearly in Fig.~{\ref{fig:aox_aouv}} where moving along the optical luminosity axis (the grey lines) results in the greatest change in wind parameters, compared to moving along the X-ray and UV axes.}

\begin{figure}
    \centering
    \includegraphics[width=\linewidth]{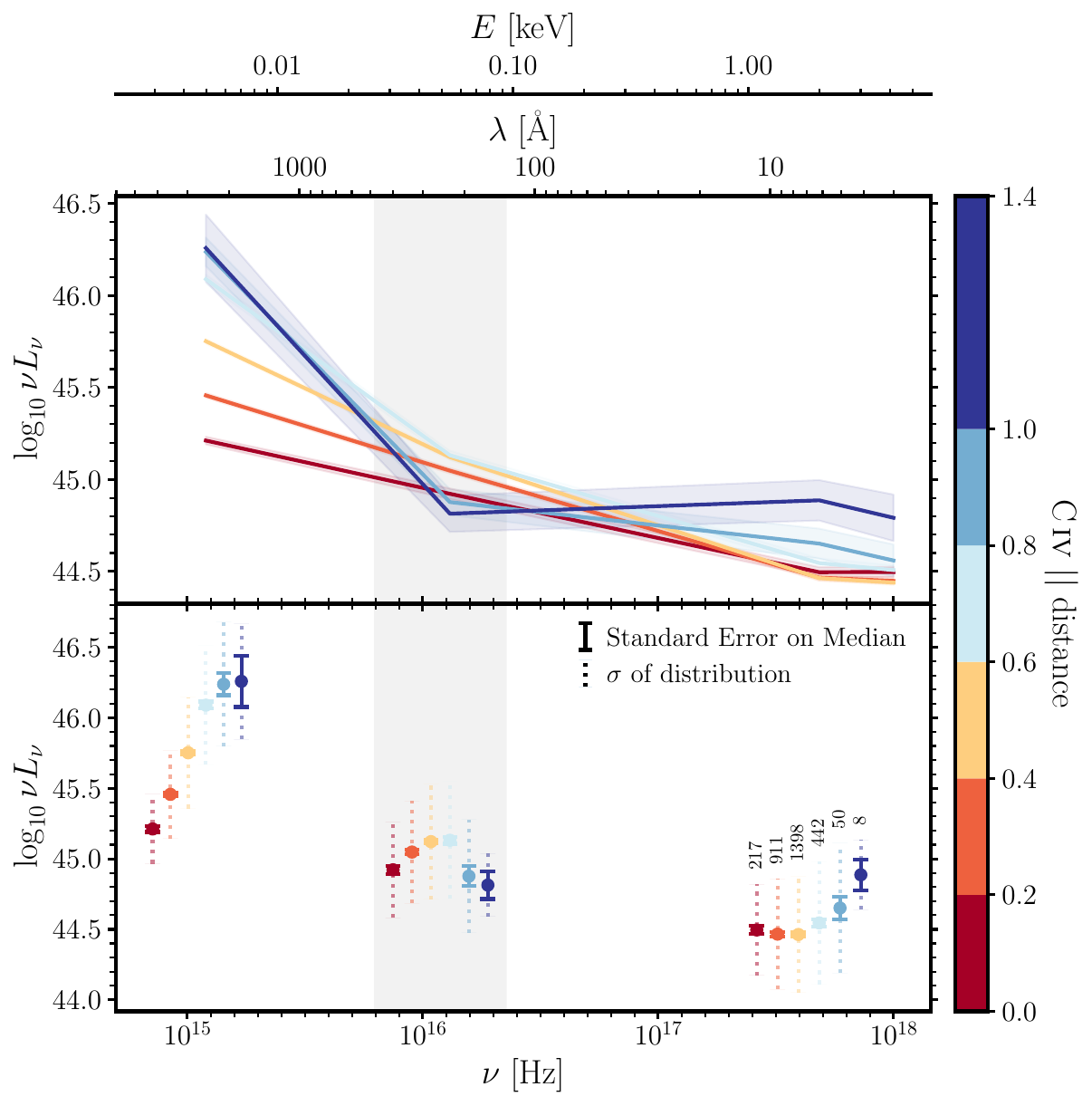}
    \caption{\hlii{Comparison of the spectral energy distributions measured at optical (left), UV (middle), and X-ray (right) wavelengths as a function of {\CIV} distance (colour bar). The grey shading behind the UV measurements is to remind the reader of the uncertainty in the calibration of the {\HeII} to UV luminosity. Top: connected SEDs with standard error on the mean as shaded regions. The median $\Gamma$ in each bin is used to extrapolate the SED from 2\,keV to higher energies. Bottom: measurements of optical, UV, and X-ray represented as points showing the distribution of values within each {\CIV} distance bin. The number of sources in each {\CIV} distance bin is presented above the X-ray points. For clarity, the points are offset along the abscissa. The optical and UV emission both increase in a manner to increase the optical-to-UV ratio as wind strength increases, but in order to drive the strongest winds, the UV emission decreases while optical luminosity continues to increase.}}
    \label{fig:SED}
\end{figure}

Figure~\ref{fig:SED} demonstrates how the three luminosities change in absolute terms in different {\CIV} distance bins.
As {\CIVdist} increases, the optical luminosity increases. The UV luminosity increases as {\CIVdist} increases until {\CIVdist} $>0.8$ ({\CIV} blueshift $\simeq 2000$\,{\kms}), at which point the UV luminosity drops to below the level in the lowest {\CIVdist} bins. The X-ray luminosity stays roughly constant until {\CIVdist} $> 0.6$ ({\CIV} blueshift $\simeq 1000$\,{\kms}) when the X-ray luminosity starts to increase. The spread of luminosities in each {\CIVdist} bin is large enough for the relationship between {\CIVdist} and UV or X-ray luminosities to be consistent with a flat relationship. In fact, the Spearman and Pearson coefficients that describe the correlations between the wind strength and the luminosities (see Table~\ref{tab:spear_pears}) reveal that most correlated with wind strength is {\aouv}, followed by $L_{2500}$ and {\aox}. All other parameters show little correlation with wind strength. From a purely photoionization point of view, lower {\aouv} is correlated with lower {\CIV} EW. The correlation is marginally stronger between {\aouv} and {\CIV} distance than the correlation between {\aouv} and {\CIV} EW which has a Spearman coefficient $\sim0.70$. However, note from Table~\ref{tab:spear_pears} that the {\CIV} blueshift alone is most strongly correlated with {\aouv} also.

\begin{table*}
    \centering
    \caption{Spearman and Pearson correlation coefficients and associated $p$-values (bracketed values) for the relationships between {\CIVdist} or {\CIV} blueshift and luminosities or spectral slopes. Spearman (Pearson) coefficients close to $\pm$1 imply monotonic (linear) relationships. Small \textit{p}-values imply that the probability of an uncorrelated dataset producing coefficients as least as strong as the measured values is low.}
    \begin{tabular}{c|cc|cc} 
& \multicolumn{2}{|c|}{{\CIV} $\vert\vert$ distance} & \multicolumn{2}{|c|}{{\CIV} blueshift} \\ 
\hline 
Variable & Spearman & Pearson & Spearman & Pearson \\ 
\hline 
$L_{2\text{keV}}$ & 0.05 (3e-03) & 0.06 (1e-03) & 0.03 (6e-02) & 0.00 (9e-01) \\ 
$\Gamma$ & 0.12 (2e-11) & 0.12 (1e-10) & 0.13 (4e-12) & 0.12 (8e-12) \\ 
$L_{2500\text{\AA}}$ & 0.57 (1e-261) & 0.58 (1e-268) & 0.41 (6e-122) & 0.40 (6e-114) \\ 
$L_{54\text{eV}}$ & -0.06 (5e-04) & 0.11 (3e-10) & -0.17 (1e-20) & 0.02 (2e-01) \\ 
$\alpha_{\text{ox}}$ & -0.55 (3e-238) & -0.61 (2e-308) & -0.40 (4e-115) & -0.46 (2e-160) \\ 
$\alpha_{\text{ouv}}$ & -0.67 (<1e-308) & -0.72 (<1e-308) & -0.59 (5e-285) & -0.56 (4e-252) \\ 
$\alpha_{\text{uvx}}$ & 0.09 (5e-07) & -0.07 (6e-05) & 0.18 (2e-23) & -0.03 (1e-01) \\ 
\hline 
\end{tabular}
    \label{tab:spear_pears}
\end{table*}

\section{Discussion}
\subsection[Comparison to prior samples probing X-ray properties in the extremes of the CIV space]{Comparison to prior samples probing X-ray properties in the extremes of the C\,{\sevensize IV} space}
\label{sec:dis_lit}

As previously discussed, the X-ray selected sample is shifted towards lower blueshifts and higher EWs (or lower {\CIVdist}s), in contrast with the optically selected sample. Here we consider our results in the wider context by comparing to samples in the literature targeting quasars at the extremes of the {\CIV} emission space. With \textit{Chandra} observations, \citet{fu_nature_2022} measured the X-ray properties for a sample of 10 quasars with large {\CIV} EWs ($>150$\,{\AA}). Meanwhile, \citet{luo_x-ray_2015} and \citet{ni_connecting_2018, ni_sensitive_2022} compiled a sample of 32 quasars with progressively deeper \textit{Chandra} observations and {\CIV} EW $<10$\,{\AA} (weak-lined quasars; WLQs). (\citealt{shlentsova_x-ray_2026}; hereafter \citetalias{shlentsova_x-ray_2026}) obtained \textit{Chandra} data for 10 optically bright quasars (2500\,{\AA} luminosities $\log_{10}(\nu L_\nu/\text{erg\,s}^{-1}) > 46.8$) with {\CIV} blueshifts~$>1400$\,{\kms}.

\begin{figure*}
    \centering
    \includegraphics[width=0.8\linewidth]{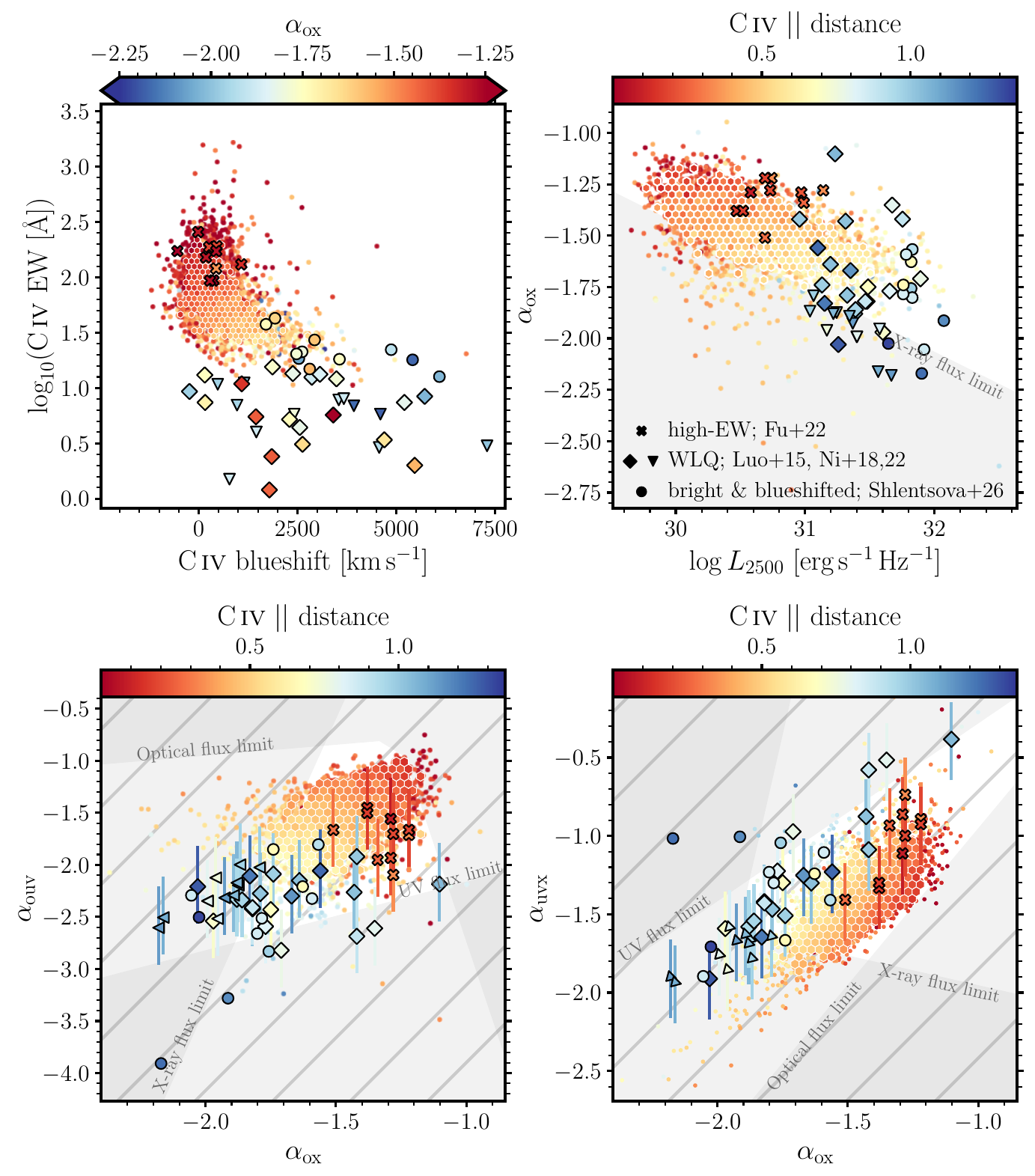}
    \caption{Comparison of our results with the high-EW sample of \citet[][crosses]{fu_nature_2022} and extending to low-EWs (and larger blueshifts) with the weak-lined quasars of \citet{luo_x-ray_2015} and \citet[][diamonds with X-ray upper limits as triangles]{ni_connecting_2018, ni_sensitive_2022}. The circles are the optically bright and highly blueshifted sample from \citet{shlentsova_x-ray_2026}. As before, our sample is presented as hexagons and dots. Top-left: {\CIV} emission space colour-coded by {\aox}. The high-EW sample agrees with our sample. The bright and blueshifted sample [and the weak-lined quasars (WLQs)] extend the explored {\CIV} space to higher blueshifts (and lower EWs) with a range of {\aox} values that are predominantly low. Top-right: {\aox} as a function of $L_{2500}$ colour-coded by {\CIVdist}. The high-EW sample objects are optically brighter than our sample at the same {\CIVdist}. The WLQs probe the bright end of our optical luminosity distribution and typically probe the low end of the {\aox} distribution. The bright and blueshifted sample extend the parameter space to larger luminosities, with {\aox} values consistent with an extrapolation of our results. Bottom-left and -right: {\aouv}, and {\auvx}, as functions of {\aox} respectively. Similar to Fig.~\ref{fig:aox_aouv} and Fig.~\ref{fig:aox_auvx} but note the change of the colour bar scales to include the larger {\CIVdist}s measured in the WLQs and bright and blueshifted sample. UV luminosity measurements and their errors are set at the 16th, 50th, and 84th percentiles of our sample's {\HeII} line luminosity distribution (see text for details). The high-EW sample is consistent with ours and the WLQs and bright and blueshifted sample extend this space to lower {\aox} and {\aouv} and higher {\auvx} where they have larger {\CIVdist} measurements as expected based on our results.}
    \label{fig:literature}
\end{figure*}

Figure~\ref{fig:literature} compares the {\CIV} and X-ray properties of our X-ray selected sample to those of the extreme {\CIV} samples. The $\Delta\alpha_\text{ox}$ values of the \citet{fu_nature_2022} sample are consistent with those of our sample where they overlap in {\CIV} space (top-left panel). Our sample probes lower optical luminosities than the \citet{fu_nature_2022} sample due to their optical magnitude cut of $i$-band magnitude $< 20$ (a resulting minimum $\log L_{2500}=30.47$), and when combined with the EW~$>150$\,{\AA} requirement leads to the sample being biased towards the lower end of our sample's range of {\CIVdist}s at their location in the $L_{2500}$--{\aox} plane (top-right panel). They also tend to be X-ray brighter than our sample (i.e., less negative {\aox} values).

We estimate {\aouv} and {\auvx} of the high-EW and WLQ samples using the optically selected sample's distribution of {\HeII} line luminosity. We take the 50th percentile of the distribution and estimate {\aouv} (with lower and upper errorbars based on the 16th and 84th percentiles). \hlii{Note, we are assuming that the WLQs and high-EW samples have the same underlying {\HeII} luminosity distribution as the optically selected sample. The large errorbars account for some of the uncertainty of the {\HeII} distributions of these samples but the line luminosities for the high-EW sample may be underestimated, and overestimated for the WLQs.}
Given the large uncertainties on the {\aouv} values, the high-EW sample have consistent wind properties with our sample at the same location in the {\aox}--{\aouv} and {\aox}--{\auvx} planes.

The WLQs, however, probe lower EWs and larger blueshifts than our sample, making them an interesting population to compare to ours (top-left panel of Fig.~\ref{fig:literature}). Note that the EW differences are partly by design as we exclude sources with {\CIV} EW $<10$\,{\AA} from our sample due to the limitations of our spectrum reconstruction method. In contrast, the bright and blueshifted sample of \citetalias{shlentsova_x-ray_2026} have {\CIV} EW $>10$\,{\AA}. The WLQs extend to lower {\aox} than our sample at the same optical luminosity (top-right panel) as one may expect given their location in {\CIV} space and the results of \citet{rivera_characterizing_2020} and \citet{timlin_correlations_2020, timliniii_what_2021}. The \citetalias{shlentsova_x-ray_2026} sample adds sources at the high-luminosity end of the $L_{2500}$--{\aox} relation with values in agreement with our sample. 

The WLQ and \citetalias{shlentsova_x-ray_2026} samples allow us to extend our analysis to larger {\CIVdist}s in the {\aox}--{\aouv} and {\aox}--{\auvx} planes (bottom panels). The \citetalias{shlentsova_x-ray_2026} objects are present in our optically selected comparison sample, thus we estimate {\aouv} and {\auvx} from the {\HeII} measurements from \citetalias{rankine_bal_2020}. The \citetalias{shlentsova_x-ray_2026} sample and WLQs, that typically exhibit stronger (faster) {\CIV} winds, are consistent with being relatively UV weak compared to their optical luminosities (i.e., have low {\aouv}). However, the {\aouv} measurements are estimates for the WLQ sample. In contrast, while these samples tend to be X-ray weaker compared to their optical luminosities (they have lower {\aox} compared to the bulk of our sample), they are not X-ray weak when compared to their (suppressed) UV luminosities, i.e., their X-ray brightness has a range but is typical given their UV brightness. The WLQ sample is, on average, optically bright relative to their X-ray luminosity, consistent with an extrapolation of our sample. While we have estimated the UV luminosity, and therefore {\aouv} and {\auvx} values, these estimates place the WLQs in similar regions of parameter space to the bright and blueshifted sample of \citetalias{shlentsova_x-ray_2026}. 
The distinguishing feature of the bright and blueshifted \citetalias{shlentsova_x-ray_2026} sample is that they are optically bright relative to their UV (and thus also X-ray), which appears key for driving strong winds.  

\subsection{Relation, or lack thereof, between X-ray and wind properties}
\label{sec:dis_xray}

\begin{figure}
    \centering
    \includegraphics[width=\linewidth]{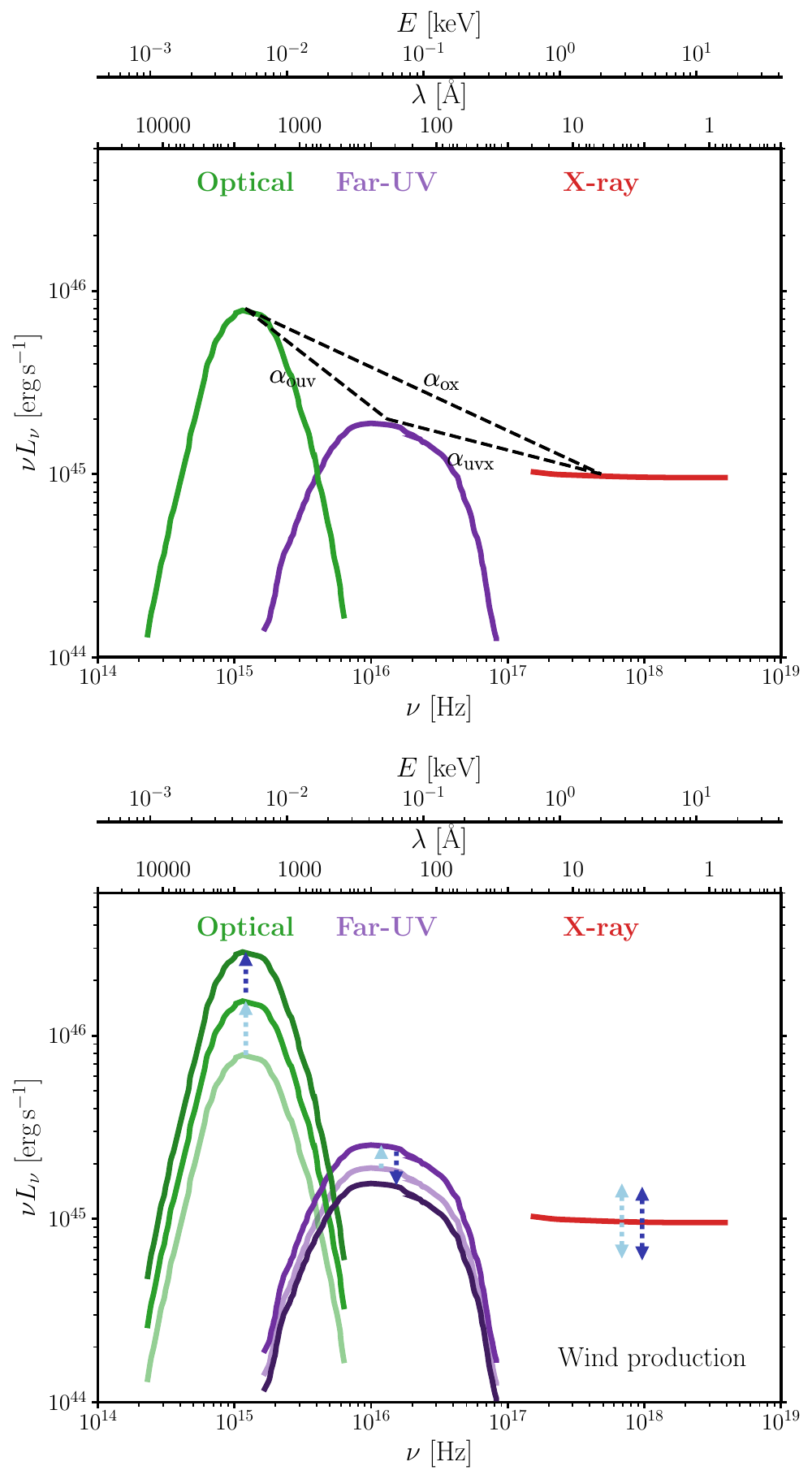}
    \caption{Sketch of the optical (green), far-UV (purple), and X-ray (red) components of the SED showing the spectral indices presented in Section~\ref{sec:wind_xray} (top). The y-axis scaling is arbitrary; only the relative magnitudes of each component are important. As shown in the bottom panel, in order to produce a wind, the ratio of the optical to UV luminosity must increase by increasing the optical emission and UV by a lesser extent whilst the X-ray component can change but need not do so (shown by the light blue arrows). To produce the fastest winds in our sample, the optical emission increases further but the UV emission decreases to the lowest in our sample (dark blue arrows).}
    \label{fig:SEDsketck}
\end{figure}

By measuring the 2500\,{\AA} continuum luminosity, the {\HeII}\,$\lambda1640$ line luminosity, and the 2\,keV X-ray luminosity, we have estimates of three points in the SED of quasars that we can track as a function of wind strength, as probed by the {\CIV} emission line. In Fig.~\ref{fig:SEDsketck}, we sketch the SED of a quasar and mark the spectral slopes used throughout this paper: {\aox}, {\aouv}, and {\auvx}. Within our quasar sample, in order to increase the wind strength, the optical luminosity must increase. The UV luminosity is also seen to increase as moderately fast winds are driven; however, to obtain the fastest velocities in our sample, the UV luminosity is, on average, lower than for the quasars with no observable winds (Figure \ref{fig:SED}). The changes in the UV are less significant than the changes in the optical (see Fig.~\ref{fig:SED}). These parameters change in such a way that the ratio of optical-to-UV luminosity is increasing as wind strength increases ({\aouv} becomes more negative). Meanwhile, the X-ray luminosity is, on average, close to constant as wind strength increases. 

\hli{On first glance, the lack of a correlation between the X-ray properties and {\CIV} blueshift, and the shift towards lower {\CIV} blueshifts for our X-ray selected sample, appear to be contradictory. These findings may be reconciled in a scenario where the ratio of optical-to-UV luminosity is the governing factor driving a wind and given that the optical, UV, and X-ray luminosities do track each other (e.g., the well-documented $L_{2500}$--{\aox} relation) the shift towards lower {\CIV} distances in our X-ray selected sample is a secondary effect. Consider the bottom-left panel of Fig.~\ref{fig:literature}: \hlii{the WLQ sources (that were extracted as a small subset from very large and wide area samples) are lying towards the bottom left of the plot, i.e., would typically be below our X-ray flux limit. However, it is not their X-ray weakness that means they have strong winds; it is that they are UV weak (low {\aouv}, they are low on the y-axis). Such low {\aouv} sources are rare; they need to be optically bright relative to the UV which thus tends to be faint. These object tend also to be X-ray faint (and below our X-ray limit). Moving in the x-axis direction has minimal impact on the wind strength as there exist WLQs with strong X-ray (far to the right of the {\aox} axis) but have winds.} Following the procedure in Section~{\ref{sec:civ}}, we match the optically selected sample to the X-ray selected sample in {\aouv}, but still recover a $p$-value~$\ll3\times10^{-7}$ for the 2-D KS test in {\CIV} space, thus rejecting the hypothesis that they are drawn from the same distribution. While there may exist a general population of quasars in the Universe, it is clear that optically selected and X-ray selected samples are biased in their own separate ways such that it is difficult to \textit{create}, from an optically selected sample, a quasar sample with similar properties to an X-ray selected sample.}

The lack of a correlation between the 2\,keV luminosity and {\CIV} properties is not entirely surprising: it is not the 2\,keV photons that are responsible for ionising the {\CIV}. The high-energy X-ray photons are few enough in number that the EUV portion of the SED has a $\sim$7\,dex greater interaction rate with the gas than the X-rays, even for a hard SED \citep[$\alpha_\text{ox}=-0.05$; see Appendix A of][]{temple_testing_2023}. However, a correlation would have suggested particular coronal properties associated with quasars which are hosting winds which is expected in some models \citep[e.g.,][]{elvis_structure_2000}, as X-ray ionisation can be a precondition for the line-driving that powers the wind; instead, the far/extreme-UV is responsible. 

As mentioned previously, the use of the {\HeII} line luminosity as a tracer for the UV luminosity relies on the assumption of a constant covering factor of optically thick {\HeII}-emitting gas. However, the structure of the material may differ between a low-blueshift and high-blueshift object, or in objects with different $\lambda_\text{Edd}$. \citet{temple_testing_2023} have shown that there is good agreement between model predictions and the data for the strength of {\HeII} ionizing radiation above $\lambda\sim0.2$; below this Eddington ratio, however, the agreement breaks down, suggesting that either the model SEDs are less accurate or the broad line region structure is changing and the assumptions are invalid. \hlii{It may also be that the far-UV portion of the SED is being filtered out as the wind reprocesses the SED }\citep[e.g.,][]{leighly_hubble_2004, hopkins_multi-phase_2024}. {Whether or not this filtered SED is the SED that we observe depends on the structure and geometry of the wind and our viewing angle.}

In addition to the above, our results are of course caveated by the fact that our sample is, by design, X-ray selected thus is biased towards X-ray bright objects. However, the bright and blueshifted sample from \citetalias{shlentsova_x-ray_2026} (whose selection was independent of their X-ray properties) follow the trends in our sample. The authors also suggest that only above {\CIV} blueshifts of 3000\,{\kms} does X-ray weakness become statistically significant, a regime not well sampled by our study, \hlii{thus, the bias against strong blueshifts in our sample may affect the derived correlations and their physical interpretation.}

X-ray emission from quasars is also known to be highly variable on all timescales \citep[e.g.,][]{timlin_long-timlescale_2020} such that any correlations between wind strength and X-ray properties may be diluted. \citet{rivera_exploring_2022} estimate errors of $\alpha_\text{ox}\sim0.19$ based on the quantification of X-ray variability by \citet{timliniii_what_2021}.
Additionally, the geometry of the wind and ionising source may result in the material in the wind being impacted by a different SED from that we observe.

\subsection{The relation between wind strength and Eddington ratio or black hole mass}
\label{sec:dis_bh}

\begin{figure*}
    \centering
    \includegraphics[width=\linewidth]{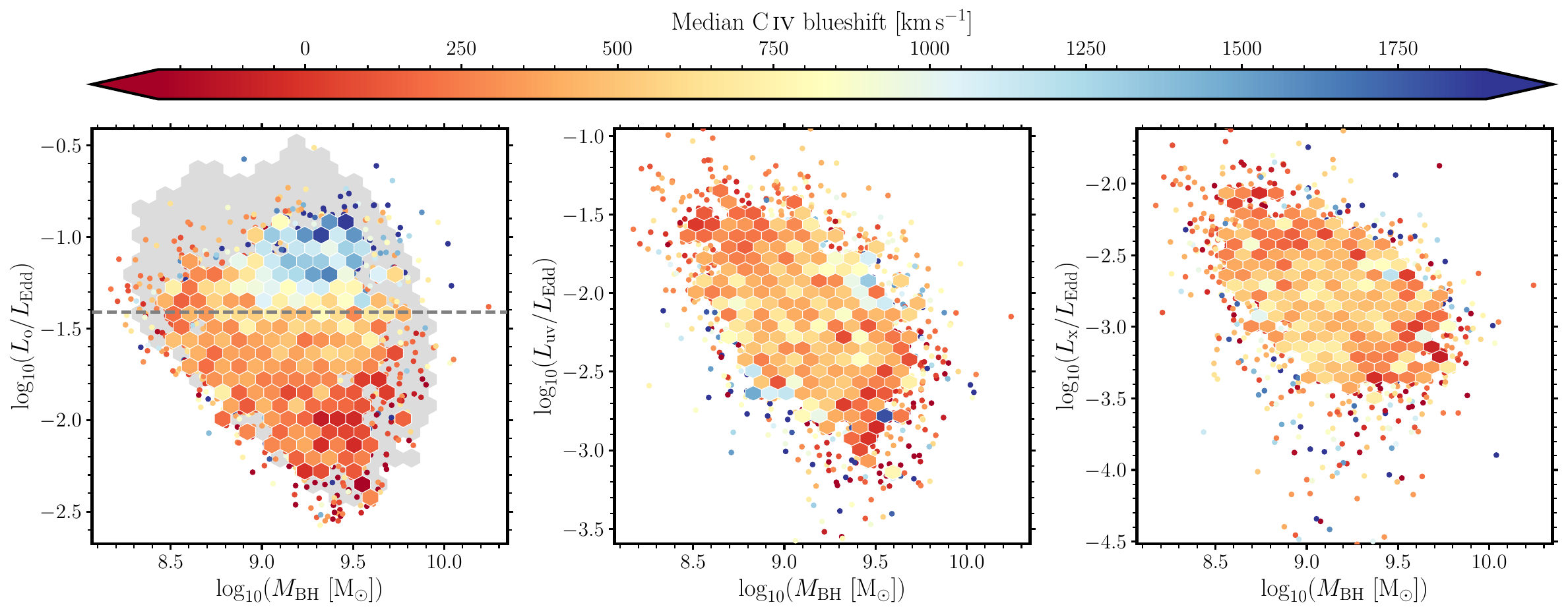}
    \includegraphics[width=\linewidth]{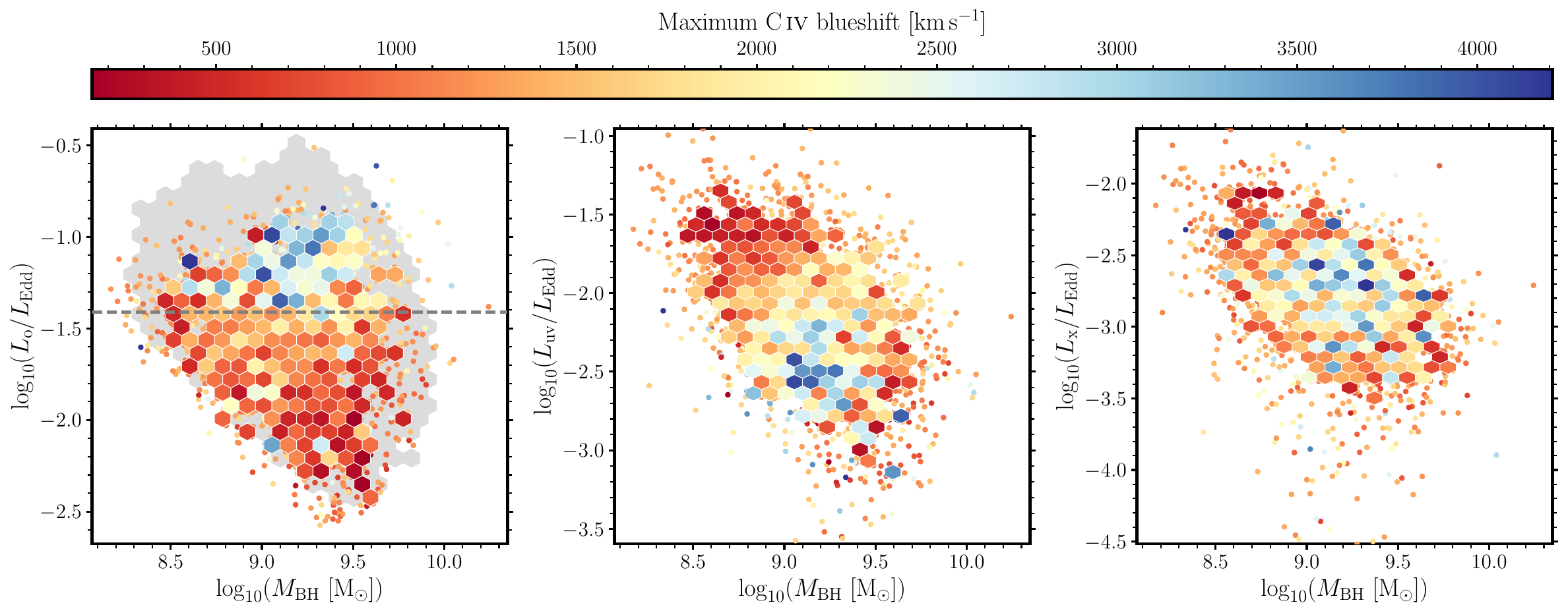}
    \caption{Top: Median {\CIV} blueshift in bins of black hole mass and optical, UV, and X-ray luminosity as a fraction of the Eddington luminosity, from left to right. The hexagons are sized such that the vertical height of a bin in all panels equates to the same range in luminosity for a given $L_\text{Edd}$. In the left-most panel, the joint distribution of $L_\text{o}/L_\text{Edd}$ and {\BHM} and the correlation with {\CIV} blueshift is similar to that in figure~3 of \citet{temple_testing_2023} using $\lambda_\text{Edd}$ since the Eddington ratio is just a rescaling of $L_\text{o}/L_\text{Edd}$ by a constant bolometric correction. The grey area indicates the optically selected sample, which extends to larger $L_\text{o}/L_\text{Edd}$ values. There are no discernible trends between {\CIV} distance and $L_\text{uv}/L_\text{Edd}$ or $L_\text{x}/L_\text{Edd}$, and the correlation with {\BHM} observed in the left-most panel is diluted in the middle and right-most panels. Bottom: Similar to the top panels but colour-coded by maximum {\CIV} blueshift in each bin. In the middle panel, the maximum {\CIV} distance observed increases as $L_\text{uv}/L_\text{Edd}$ decreases.}
    \label{fig:mbh_lledd}
\end{figure*}

Winds are expected to be driven in high-Eddington ratio sources, and there is a dependence on black hole mass \citep[e.g.,][]{temple_testing_2023}. Our results have shown that the ratio of optical-to-UV emission is an important factor for driving winds. Thus, we consider $L_\text{o}$ and $L_\text{uv}$ scaled by the Eddington luminosity (and $L_\text{x}$ for completeness). Similarly to figure~3 of \citet{temple_testing_2023}, the left-most panel of Fig.~\ref{fig:mbh_lledd} displays the $L_{2500}$ scaled by $L_\text{Edd}$ against black hole mass and colour-coded by {\CIV} blueshift. The grey hexagons show the location of the optically selected sample which extends to higher $L_\text{o}/L_\text{Edd}$ than our X-ray selected sample. The bolometric luminosity used in \citet{temple_testing_2023} to determine the Eddington ratio is defined by scaling the 3000\,{\AA} monochromatic luminosity by a constant bolometric correction of 5.15 \citep{richards_spectral_2006, shen_catalog_2011}; although the authors investigate the effect of a varying bolometric correction. The {\CIV} blueshift follows the same trends as seen in figure~3 of \citet{temple_testing_2023}, with faster winds present in higher $L_\text{o}/L_\text{Edd}$ sources. The dashed grey line indicates the rescaled $\lambda_\text{Edd}\sim0.2$ suggested by the authors as the boundary above which winds are possible. We also observe a similar {\BHM} dependence whereby wind strength increases as {\BHM} increases (above $\lambda_\text{Edd}\sim0.2$). Our sample, however, misses the centre of the `wedge' in that the highest blueshift objects are excluded. The middle and right-hand panels show that the UV and X-ray luminosities scaled by $L_\text{Edd}$ show no trend with median {\CIV} blueshift. However, if instead we colour-code by maximum {\CIV} blueshift in each bin (bottom panels), there is still no correlation with $L_\text{x}/L_\text{Edd}$, but there appears to be a threshold of $L_\text{uv}/L_\text{Edd}\sim0.01$ above which wind launching is suppressed.

\begin{figure}
    \centering
    \includegraphics[width=\linewidth]{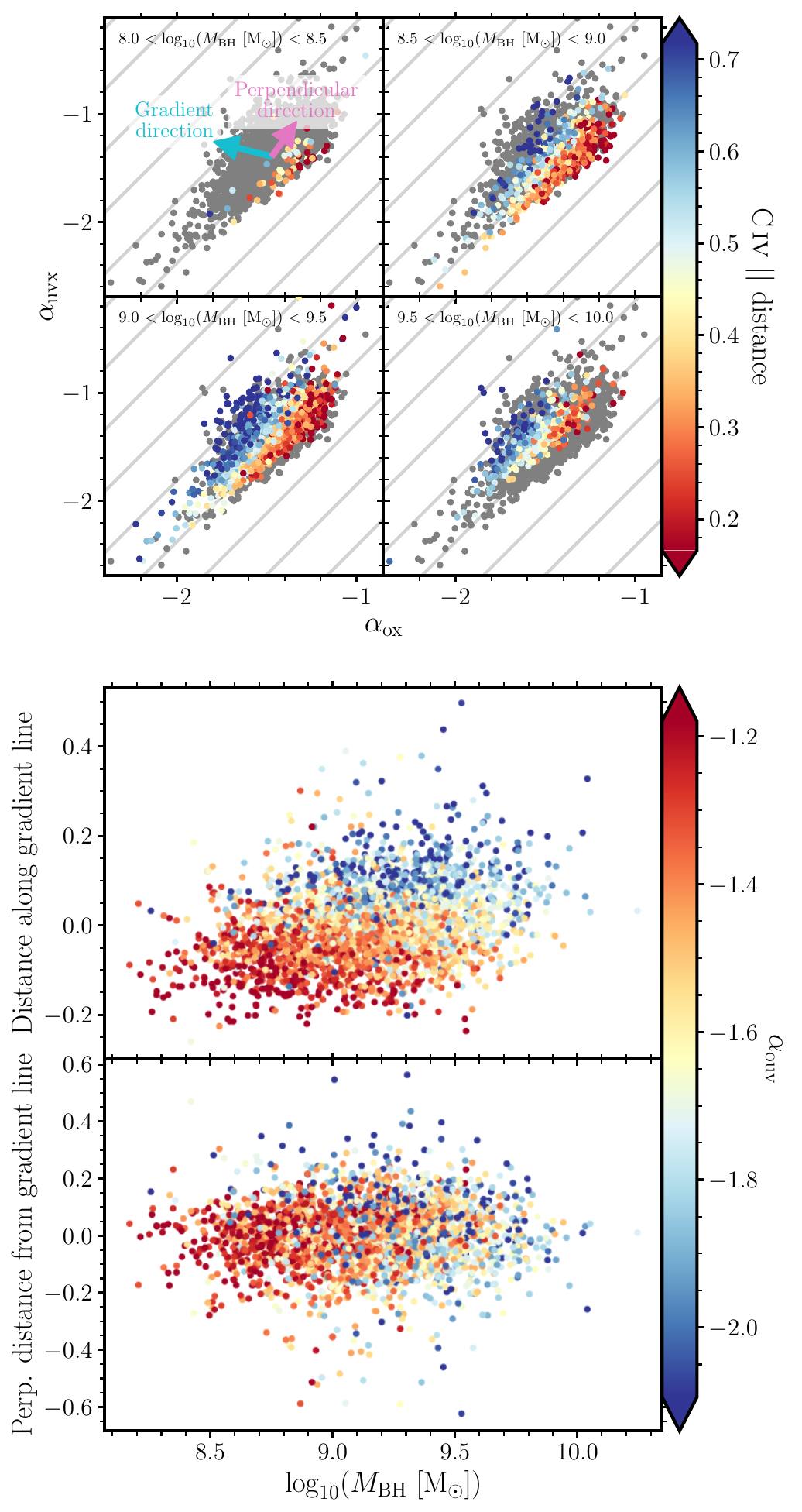}
    \caption{\hlii{The {\aox}-{\auvx} space colour-coded by {\CIV} distance (as in Fig.~{\ref{fig:aox_auvx}}), binned by {\BHM} across the four top panels. The objects outside each {\BHM} bin are shown as grey points. The cyan arrow marks the gradient vector of {\CIV} distance in {\aox}-{\auvx} space ($\nabla\CIV$), i.e., the direction of greatest change of {\CIV} distance, evaluated at the mean of ({\aox},{\auvx}) of the whole sample. There is a mild but significant correlation between {\BHM} and distance along the gradient line. The pink arrow is perpendicular to the gradient vector. The middle panel reveals this mild dependence directly where we plot the distance along the gradient line against {\BHM}, points colour-coded by {\aouv}. The bottom panel contains the distance perpendicular from the gradient line (along the pink arrow, top panel) against {\BHM}. {\BHM} is not responsible for the spread in objects perpendicularly from the line. The strong correlation between {\aouv} and distance along the gradient line, along with the lack of correlation with the perpendicular distance from $\nabla\CIV$, indicates that $\nabla\alpha_\text{ouv}$ and $\nabla\CIV$ are close to parallel, i.e., changes in {\CIV} distance are strongly driven by changes in {\aouv}.}}
    \label{fig:mbh_dist}
\end{figure}

For sources with $\log_{10}(L_\text{o}/L_\text{Edd}) > -1.4$ (equivalent to $\lambda_\text{Edd}\gtrsim0.2$), wind strength increases as {\BHM} increases; however, from Figs.~\ref{fig:aox_auvx} and \ref{fig:aox_aouv} it is clear that the ratio of optical-to-UV emission is important for affecting wind strength. \hlii{We investigate the impact of {\BHM} on our results by first splitting the sample into {\BHM} bins of width 1 dex and plotting these subsamples in {\aox}-{\auvx} space (top panels of Fig.~{\ref{fig:mbh_dist}}). In the majority of the mass bins, there exist objects with a range of {\CIV} profiles, and {\aox} and {\auvx} values. To quantify the {\BHM} dependence on {\CIV} distance we consider how {\BHM} changes in our sample as we move in the direction of greatest change of {\CIV} distance from the mean in {\aox}-{\auvx} space (denoted $\nabla\CIV$; see cyan arrow in the top panel of Fig.~{\ref{fig:mbh_dist}}). 
The parameter $\nabla\CIV$ is estimated by assuming that {\CIV} distance forms a plane in (\aox,\,\auvx,\,{\CIV}) given by ${\CIV}\ \text{distance} = A\alpha_\text{ox} + B\alpha_\text{uvx} + C$. Fitting for $A$, $B$, and $C$, we find $\nabla{\CIV} = (A\approx-1.32, B\approx0.57)$ and we construct the gradient line starting from the mean values of {\aox} and {\auvx} (cyan arrow in the top-left panel of Fig.~{\ref{fig:mbh_dist}}). In a similar manner to the {\CIV} distance, we project each point in {\aox}-{\auvx} space onto the straight line that extends along the gradient vector, and calculate the distance along this line with positive values in the direction of the cyan arrow. We calculate the distance along this line. Additionally, we measure the perpendicular distance from this line, with positive values assigned to points above the line (in the direction of the pink arrow in the top-left panel of Fig.~{\ref{fig:mbh_dist}}). The bottom two panels of Fig.~{\ref{fig:mbh_dist}} presents these parameters as a function of {\BHM}. $\nabla\CIV$ is clearly close to parallel with $\nabla\alpha_\text{ouv}$ as observed by the correlation between distance along $\nabla\CIV$ and {\aouv}, and the lack of any correlation between {\aouv} and the perpendicular distance from $\nabla\CIV$. There is a weak but significant correlation between {\BHM} and distance along the $\nabla\CIV$ line (Spearman rank coefficient $\sim$ 0.31; $p$-value~$\simeq$~3.38e-70; middle panel) suggesting that {\BHM} is responsible for some of the width of the distribution of {\aouv}.}
Of course, movement in the {\CIV} space of an individual object (which \citealt{rivera_characterizing_2020} explored for a sample of reverberation-mapped quasars) would not mean a reduction in {\BHM}.
The correlation between {\aouv} and {\BHM} is also observed in the bottom panel of Fig.~\ref{fig:mbh_dist}, but there is no dependence on the perpendicular distance from the $\nabla\CIV$ line.
Our findings that increasing $L_\text{o}$ while keeping $L_\text{uv}$ relatively low is key to driving a wind explains why higher mass BHs exhibit stronger winds, assuming a decreasing inner disc temperature (and thus less UV emission) as {\BHM} increases as expected in a geometrically thin, optically thick standard \citet{shakura_black_1973} disc.

\section{Conclusions}
\label{sec:conc}
We have examined the accretion disc wind properties -- as probed by the {\CIV} emission line -- of the SDSS-V X-ray selected quasar sample. Our sample contains 3027 objects, with redshifts $1.5 \le z \le 3.5$, robust optical reconstructions (following the methodology of \citetalias{rankine_bal_2020}), and \textit{eROSITA} X-ray spectral analysis (including eFEDS sources from \citealt{liu_erosita_2022} and extending to eRASS). We remeasure redshifts for more accurate {\CIV} emission line measurements. Our main findings are as follows:

\begin{enumerate}
    \item We compared the {\CIV} properties of our X-ray selected sample with the optically-selected sample from \citetalias{rankine_bal_2020}. The X-ray selection misses the highest blueshift and lowest EW sources (higher {\CIV} distances). See Figure~\ref{fig:CIV45}.
    \item To explain the shift in {\CIV} space, we examined the differing redshift distributions, optical luminosity distributions, \textit{eROSITA}'s bias towards softer X-ray sources, \textit{eROSITA} sensitivity, and optical flux threshold. None of the selection biases alone can explain the occupation of {\CIV} space (see Section~\ref{sec:civ}).
    \item The optical luminosity drives the correlation between {\aox} and {\CIV} space (see Figs.~\ref{fig:CIV_L} and \ref{fig:CIV_HeII}). 
    \item The X-ray properties -- namely {\Lx}, and spectral slope -- are not strongly correlated with {\CIV} parameters (see Fig.~\ref{fig:CIV_xray}).
    \item The ratio of optical-to-UV emission ({\aouv}) correlates the strongest with {\CIV} emission, followed by optical luminosity and {\aox}, see Figs.~\ref{fig:aox_auvx} and \ref{fig:aox_aouv} as well as Table~\ref{tab:spear_pears}. 
    \item Figure~\ref{fig:SED} demonstrates that the optical and UV emission both increase in a manner to increase the optical-to-UV ratio as wind strength increases, but to drive the strongest winds, the UV emission decreases while optical luminosity continues to increase.
    \item We compare our results to samples in the literature in Fig.~\ref{fig:literature}. Our results are consistent with the high-EW sample from \citet{fu_nature_2022}. Similarly, the weak-lined quasar samples \citep{luo_x-ray_2015, ni_connecting_2018, ni_sensitive_2022}, with deeper X-ray data and weaker {\CIV} emission features than considered in our sample, follow the same trends as observed in our sample. 
\end{enumerate}
\hli{Our results are consistent
with a radiation line-driven wind whereby the ionising far-UV photons must not over-ionise the gas. Meanwhile, the hard X-ray
photons are few enough in number to have a negligible effect on the ionisation state of the material. We suggest that the shift towards larger EWs and lower blueshifts is a secondary effect caused by the correlation between the optical, UV, and X-ray components of the SEDs.}

\hlii{Direct observational access to the far-UV regime is not possible. Consequently, {\aox} is commonly employed as a proxy for the ionizing spectral shape. Within this framework, an excessively strong X-ray component indicates an over-ionization of the gas, thereby suppressing or completely inhibiting the formation of winds. However, this picture is typically based on a two-component model (comprising the accretion disc and the X-ray corona), such that the relevant spectral energy distribution is approximated by the straight line connecting these two components. Our results indicate that a three-component model, which additionally incorporates the warm disc component proposed by }\citet{kubota_physical_2018} \hlii{ that may independently be weak or strong, provides a more physically realistic description. However, the qualitative interpretation remains unchanged: the strength of the wind is governed by the balance between photons capable of exerting radiative driving and photons that ionize the gas by stripping electrons.}

Future studies involving \textsc{cloudy} \citep{cloudy_2025} simulations or by estimating the far-UV SED from rest-frame optical spectroscopy by measuring the {\HeII}\,$\lambda4686$/H$\beta$ ratio (e.g., \citealt{penston_spectrophotometry_1978, short_delayed_2023}) would improve the investigation of the relative importance of optical, UV, and X-ray emission in driving strong winds. 
Additionally, future data releases from SDSS-V including additional optical follow-up of \textit{eROSITA}-detected quasars will increase the size of the quasar sample for which we have X-ray and optical information by an order of magnitude such that X-ray stacking in bins of {\CIV} blueshift will be a meaningful exercise in order to increase the S/N of the X-ray data.
Deep targeted \textit{XMM-Newton} observations of a sub sample, and in the future, samples from NewAthena would also improve the X-ray analysis.

\section*{Acknowledgements}

We thank former University of Edinburgh undergraduate students Batrisyia Zainul and Blythe Fernandes for their contributions to this work. ALR thanks Gordon Richards for helpful conversations. The authors thank the anonymous reviewer for helpful comments which improved the manuscript.
ALR, JA, and PH acknowledge support from a UKRI Future Leaders Fellowship (grant codes: MR/T020989/1 and MR/Y019539/1). ALR also acknowledges support from a Leverhulme Early Career Fellowship (ECF-2024-186).
RJA was supported by FONDECYT grant number 1231718 and by the ANID BASAL project FB210003.
WNB acknowledges support from NSF grant AST-2407089, and the Penn State Eberly Endowment.
MC has received funding from the Hellenic Foundation for Research and Innovation (HFRI) project ``4MOVE-U'' grant agreement 2688, which is part of the programme ``2nd Call for HFRI Research Projects to support Faculty Members and Researchers''.
YD acknowledges financial support from a FONDECYT postdoctoral fellowship (3230310).
MJT acknowledges funding from a FONDECYT Postdoctoral fellowship (3220516) and from UKRI STFC (ST/X001075/1).
FEB acknowledges support from ANID-Chile BASAL CATA FB210003, FONDECYT Regular 1241005, and Millennium Science Initiative AIM23-0001.

Funding for the Sloan Digital Sky Survey V has been provided by the Alfred P. Sloan Foundation, the Heising-Simons Foundation, the National Science Foundation, and the Participating Institutions. SDSS acknowledges support and resources from the Center for High-Performance Computing at the University of Utah. SDSS telescopes are located at Apache Point Observatory, funded by the Astrophysical Research Consortium and operated by New Mexico State University, and at Las Campanas Observatory, operated by the Carnegie Institution for Science. The SDSS web site is \url{www.sdss.org}.

SDSS is managed by the Astrophysical Research Consortium for the Participating Institutions of the SDSS Collaboration, including Caltech, The Carnegie Institution for Science, Chilean National Time Allocation Committee (CNTAC) ratified researchers, The Flatiron Institute, the Gotham Participation Group, Harvard University, Heidelberg University, The Johns Hopkins University, L’Ecole polytechnique f\'{e}d\'{e}rale de Lausanne (EPFL), Leibniz-Institut f\"{u}r Astrophysik Potsdam (AIP), Max-Planck-Institut f\"{u}r Astronomie (MPIA Heidelberg), Max-Planck-Institut f\"{u}r Extraterrestrische Physik (MPE), Nanjing University, National Astronomical Observatories of China (NAOC), New Mexico State University, The Ohio State University, Pennsylvania State University, Smithsonian Astrophysical Observatory, Space Telescope Science Institute (STScI), the Stellar Astrophysics Participation Group, Universidad Nacional Aut\'{o}noma de M\'{e}xico, University of Arizona, University of Colorado Boulder, University of Illinois at Urbana-Champaign, University of Toronto, University of Utah, University of Virginia, Yale University, and Yunnan University.

This work is based on data from \textit{eROSITA}, the soft X-ray instrument aboard SRG, a joint Russian-German science mission supported by the Russian Space Agency (Roskosmos), in the interests of the Russian Academy of Sciences represented by its Space Research Institute (IKI), and the Deutsches Zentrum f\"{u}r Luft- und Raumfahrt (DLR). The SRG spacecraft was built by Lavochkin Association (NPOL) and its subcontractors, and is operated by NPOL with support from the Max Planck Institute for Extraterrestrial Physics (MPE). The development and construction of the \textit{eROSITA} X-ray instrument was led by MPE, with contributions from the Dr. Karl Remeis Observatory Bamberg \& ECAP (FAU Erlangen-Nuernberg), the University of Hamburg Observatory, the Leibniz Institute for Astrophysics Potsdam (AIP), and the Institute for Astronomy and Astrophysics of the University of T\"{u}bingen, with the support of DLR and the Max Planck Society. The Argelander Institute for Astronomy of the University of Bonn and the Ludwig Maximilians Universit\"{a}t Munich also participated in the science preparation for \textit{eROSITA}. The eROSITA data shown here were processed using the eSASS/NRTA software system developed by the German eROSITA consortium.

This research has made use of NASA's Astrophysics Data System and adstex (\url{https://github.com/yymao/adstex}).

For the purpose of open access, the authors have applied a Creative Commons Attribution (CC BY) licence to any Author Accepted Manuscript version arising from this submission.

\section*{Data Availability}

The SDSS-V data, including optical counterpart identifications to the X-ray sources as part of targeting and the optical spectra that underlie this article, will be publicly available in July 2026, at the end of the proprietary period at \url{https://www.sdss.org/}. \textit{eROSITA} data is available at \url{https://erosita.mpe.mpg.de}. After the proprietary period, data products (e.g., {\CIV} emission line measurements, X-ray spectral parameters) will be made available upon reasonable request to the corresponding author.



\bibliographystyle{mnras}
\bibliography{main.bib}



\appendix

\section{Matching SDSS targets to the \textit{eROSITA} catalogues}
\label{app:xray}

\begin{figure}
     \centering
     \includegraphics[width=\linewidth]{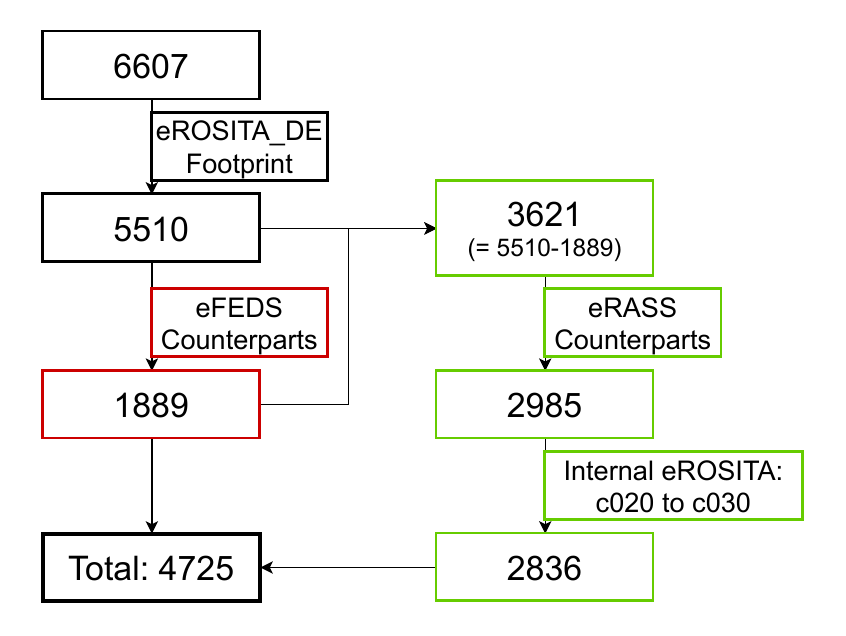}
     \caption{A schematic representation of the results of the different stages of matching the SDSS target sample to the available eROSITA data. The values in each field represent the number of targets in the selected sample at that point. Generally, the matching consisted of two stages: first matching to eFEDS (marked in red) and then matching the full remainder to eRASS:4/5 (marked in green).}
     \label{fig:app_target_match}
 \end{figure}

The sample of SDSS targets, defined in Section~\ref{sec:opt}, is matched to \textit{eROSITA} target catalogues, in order to identify the targets for which extracted X-ray spectra are available. A schematic overview of this matching is shown in Figure~\ref{fig:app_target_match}. It is important to note that although the SDSS-V targeting for the sample was in part based on eROSITA data, this targeting data differs from the more recently produced data products that we use in this work. Updates to the \textit{eROSITA} Science Analysis Software System \citep[eSASS][]{Brunner_2022} mean that \textit{eROSITA}'s unique identifiers (DETUIDs) can not be matched among different catalogues. We therefore make use of counterpart-matched catalogues. The counterpart matching is based on the Bayesian multi-wavelength counterpart search presented in \citet{Salvato_2018,Salvato_2022,Salvato_2025}.

To ensure that we make use of the deepest available observations, the matching of SDSS targets to \textit{eROSITA} objects is performed in two stages. We start from a total of 5510 unique targets present in the hemisphere of the sky accessible to the German \textit{eROSITA} consortium (eROSITA\_De). The targets are first matched to the catalogue for the \textit{eROSITA} Equatorial Final Depth Survey (eFEDS), which covered only a small fraction, $\sim$140~deg$^2$, of the sky. These are the deepest \textit{eROSITA} observations available for our targets. For eFEDS we make use of the published counterpart catalogue \citep{Salvato_2022}, which provides \textit{eROSITA} DETUIDs that can be directly linked to the spectra produced by the eSASS pipeline. We require the SDSS fibre coordinates to lie within 1 arcsecond of the optical coordinates of the counterparts. Within the eFEDS footprint, there are 1920 of our targets, 1889 of which (96 per cent) can be matched to the eROSITA catalogue data. 

Following the eFEDS matching, there are 3621 remaining unmatched targets in the eROSITA\_De sky. We attempt to match all these targets to the deepest available data from the \textit{eROSITA} All-Sky Survey (eRASS). This matching needs to be broken down into two parts. First we match to an NWAY-matched counterpart catalogue, which is only available internally to the eROSITA\_De consortium. Second, the X-ray coordinates of the identified \textit{eROSITA} counterparts need to be matched to our latest, and most up to date eRASS catalogue, which provides the DETUIDs required to identify the extracted spectra. Again, this is due to differences in the eSASS reduction used to produce the different catalogues: c020 for the counterpart catalogue and c030 for the catalogue we need. For this matching, starting from 3621 targets, we find 2985 (82 per cent) targets matching optical coordinates (also using a 1 arcsecond radius), and subsequently match 2836 of these targets (95 per cent) when matching to the most up to date eRASS catalogue (using a 10 arcsecond radius). This results in our total matched sample of 4725 targets.

\section[(z, L2500)-matched optically selected sample]{($\lowercase{{z}}$, ${L_{2500}}$)-matched optically selected sample}
\label{app:zLmatch}
\begin{figure}
    \centering
    \includegraphics[width=\linewidth]{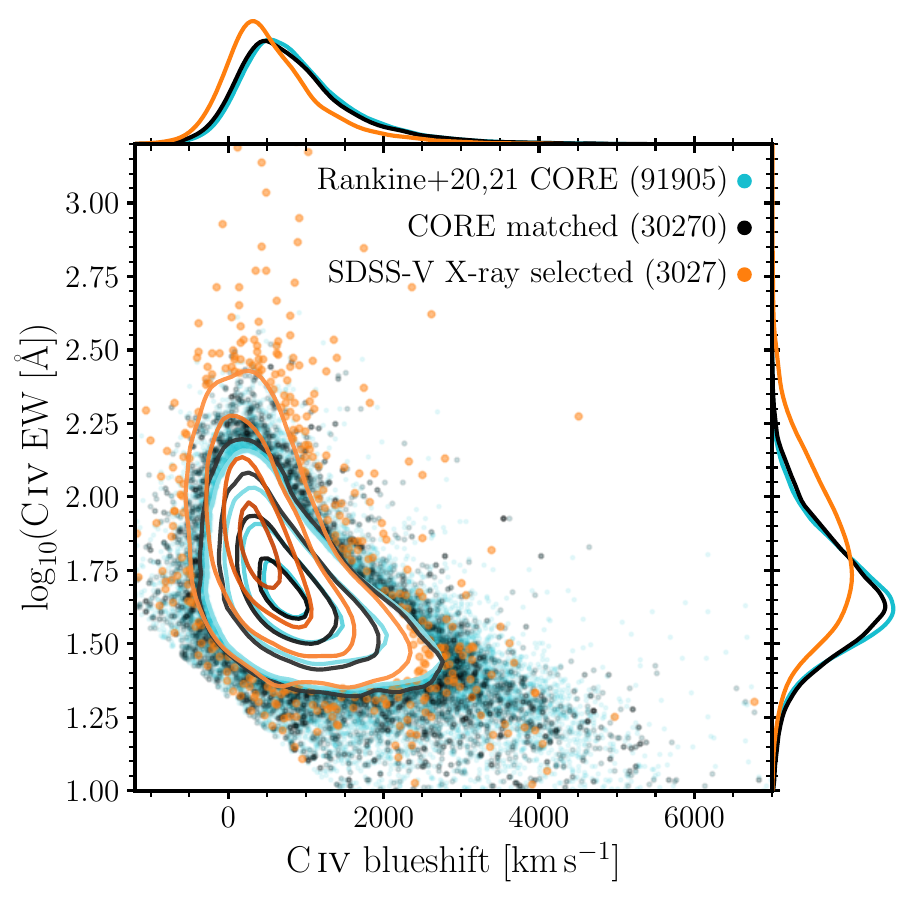}
    \caption{Similar to Fig.~\ref{fig:CIV45}, with the CORE optically selected in cyan and the X-ray selected sample in orange. The ($z$, $L_{2500}$)-matched optically selected sample is presented in black. Matching on redshift and $L_{2500}$ does not statistically reduce the discrepancy between the X-ray selected and optically selected samples.}
    \label{fig:CIV_zLmatch}
\end{figure}

Figure~\ref{fig:CIV_zLmatch} displays the ($z$, $L_{2500}$)-matched optically selected sample in {\CIV} space. The matching does not significantly shift the optically selected sample towards lower blueshifts or higher EWs. The overlapping $L_{2500}$ distributions (see Fig.~\ref{fig:lum}) as well as the broad scatter in the relationship between $L_{2500}$ and {\CIV} properties (despite the trend of increasing \textit{median} optical luminosity with increasing blueshift) are such that the matching in optical luminosity does not increase the similarities between the X-ray selected and matched optically selected samples. 

\section{Mock X-ray for SDSS-IV sample}
\label{app:mock_4}

In Section~\ref{sec:civ} we ruled out redshift and optical luminosity as the direct cause of the disparities in {\CIV} emission properties. Nevertheless, we perform the following exercise using the whole optically selected sample, and separately the redshift- and $L_{2500}$-matched subsample. We share the results of the latter here; there are no significant differences when using the whole sample.

\begin{figure}
    \centering
    \includegraphics[width=\linewidth]{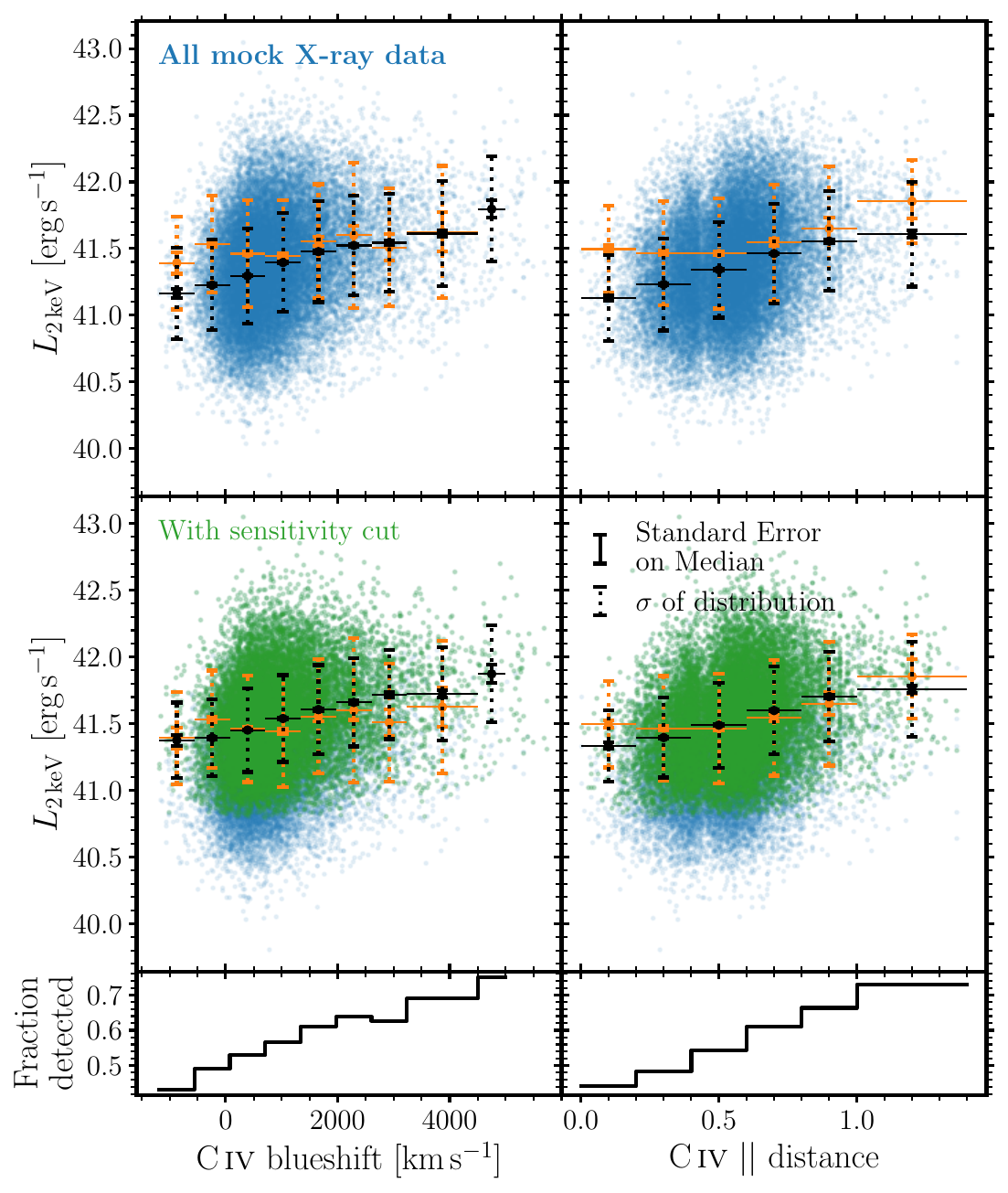}
    \caption{Same as Fig.~\ref{fig:CIV_xray} with SDSS-V median values in orange, but with the mock SDSS-IV measurements plotted as the blue points and the median {\Lx} in bins of {\CIV} metric as the black points. Top panels show the mock sample before applying the \textit{eROSITA} sensitivity limit. There is a decrease in {\Lx} towards lower blueshifts/distances that is not observed in the X-ray selected sample. Middle panels: green points are the mock measurements that satisfy the \textit{eROSITA} sensitivity limit. Bottom panels: the fraction of sources detected above the \textit{eROSITA} flux limits in bins of {\CIV} metric. Applying 
    the flux cut removes low luminosity sources primarily at low {\CIV} blueshifts/distances. }
    \label{fig:civ_xray_mock}
\end{figure}

We estimate mock \textit{eROSITA} flux measurements for the optically selected sample by first drawing mock {\aox} values from a normal distribution with mean and standard deviation based on the $L_{2500}$--{\aox} relation and intrinsic scatter from \citet{timliniii_what_2021}. We convert {\aox} to {\Lx} with equation~\ref{eq:aox} and then to an observed 0.2--2.3\,keV band flux assuming $\Gamma=2$. The top panels of Fig.~\ref{fig:civ_xray_mock} show the mock {\Lx} against {\CIV} blueshift (left) and distance (right). Compared to the SDSS-V sample (orange points) there is a visible but weak trend in the mock sample of increasing {\Lx} with increasing {\CIV} blueshift/distance (black points) which is a result of the correlation between optical luminosity and {\CIV} blueshift/distance.

We randomly apply the eFEDS flux limit to 60 per cent of the mock sample, and the eRASS limit to the remaining 40 per cent in order to match the ratio of eFEDS:eRASS sources in the X-ray selected sample. The bottom panels of Fig.~\ref{fig:civ_xray_mock} reveal the effect of the \textit{eROSITA} flux limits on the mock sample: the {\CIV} blueshift/distance--{\Lx} trend is reduced as the low-luminosity sources are removed. The objects that fall below the flux limit are primarily at low {\CIV} blueshifts, thus the \textit{eROSITA} sensitivity cannot explain the lack of high blueshift objects in our sample. 

The mock X-ray measurements follow the $L_{2500}$--{\aox} relation by design, thus neglecting X-ray weak objects, i.e., those that fall significantly below the relation. Previous studies have shown that there is a correlation between X-ray weakness and {\CIV} blueshift/distance, with large-{\CIV} distance objects more likely X-ray weak \citep[e.g.,][]{timlin_correlations_2020, rivera_exploring_2022, hiremath_x-ray_2025}. Therefore, the X-ray selection may be missing the X-ray weakest objects at the highest blueshifts.

Additionally, $\Gamma=2$, while consistent with the average $\Gamma$ of our \textit{eROSITA} sample, is somewhat softer than that found from other X-ray observatories/surveys \citep[$\Gamma=1.9$; e.g.,][]{nandra_xmmnewton_2007, peca_cosmic_2023}. Performing the test with $\Gamma=1.7$ produces no significant differences.

\vspace{4pt}
{\small\it
\noindent$^{1}$Institute for Astronomy, University of Edinburgh, Royal Observatory, Blackford Hill, Edinburgh EH9 3HJ, UK\\
$^{2}$Leibniz-Institut für Astrophysik Potsdam, An der Sternwarte 16, 14482 Potsdam, Germany\\
$^{3}$Institute of Astronomy, University of Cambridge, Madingley
Road, Cambridge, CB3 0HA, United Kingdom\\
$^{4}$Department of Astronomy, University of Washington, Box 351580, Seattle, WA 98195, USA\\
$^{5}$Instituto de Estudios Astrof\'isicos, Facultad de Ingenier\'ia y Ciencias, Universidad Diego Portales, Ej\'ercito Libertador 441, Chile\\
$^{6}$Instituto de Alta Investigaci{\'{o}}n, Universidad de Tarapac{\'{a}}, Casilla 7D, Arica, 1010000, Chile\\
$^{7}$Department of Astronomy and Astrophysics, The Pennsylvania State University, University Park, PA 16802, USA\\ 
$^{8}$Institute for Gravitation and the Cosmos, The Pennsylvania State University, University Park, PA 16802, USA\\ 
$^{9}$Department of Physics, 104 Davey Laboratory, The Pennsylvania State University, University Park, PA 16802, USA\\ 
$^{10}$Dipartimento di Fisica e Astronomia ``Augusto Righi'', Alma Mater Studiorum - Universit\`a di Bologna, Via Gobetti 93/2, 40129 Bologna, Italy\\
$^{11}$INAF - Osservatorio di Astrofisica e Scienza dello Spazio via Gobetti 93/3, 40129 Bologna, Italy\\
$^{12}$INAF - Osservatorio Astronomico di Roma, Via di Frascati 33, 00078, Monte Porzio Catone, Rome, Italy\\
$^{13}$Max Planck Institute for Extraterrestrial Physics, Giessenbachstrasse, 85748 Garching, Germany\\
$^{14}$Institute for Astronomy \& Astrophysics, National Observatory of Athens, V. Paulou \& I. Metaxa, 11532, Greece\\
$^{15}$Departamento de F\'isica, Universidad T\'ecnica Federico Santa Mar\'ia, Vicu\~na Mackenna 3939, San Joaqu\'in, Santiago, Chile\\
$^{16}$Department of Physics and Astronomy, York University, 4700 Keele St., Toronto, ON M3J 1P3, Canada\\
$^{17}$Space Telescope Science Institute, 3700 San Martin Drive, Baltimore, MD 21218, USA\\
$^{18}$Department of Astronomy, University of Science and Technology of China, Hefei 230026, People’s Republic of China\\
$^{19}$School of Astronomy and Space Science, University of Science and Technology of China, Hefei 230026, People’s Republic of China\\
$^{20}$Department of Astronomy, University of Illinois at Urbana-Champaign, Urbana, IL 61801, USA\\
$^{21}$Instituto de Astronomıa, Universidad Nacional Aut\'onoma de M\'exico A.P. 70-264, 04510, Mexico, D.F., M\'exico\\
$^{22}$Physical Sciences Division, School of STEM, University of Washington Bothell, Bothell WA, 98011, USA\\
$^{23}$National Center for Supercomputing Applications, University of Illinois at Urbana-Champaign, Urbana, IL 61801, USA\\
$^{24}$Centre for Extragalactic Astronomy, Department of Physics, Durham University, South Road, Durham DH1 3LE, UK\\
$^{25}$Institut fur theoretische Astrophysik, Zentrum fur Astronomie der Universitat Heidelberg, Albert-Ueberle-Str. 2, D-69120 Heidelberg, Germany\\
}


\bsp	
\label{lastpage}
\end{document}